\documentclass[journal, onecolumn]{IEEEtran} %,onecolumn

\usepackage{amsmath,amsthm,amssymb}
\usepackage{subfigure}
\usepackage{graphicx}

\newtheorem{Lemma}{Lemma}
\newtheorem{Theorem}{Theorem}

\newtheorem{Proposition}{Proposition}
\newtheorem{Conjecture}{Conjecture}
\newtheorem*{Conjecture*}{Conjecture}
\newtheorem{Definition}{Definition}
\newtheorem{Corollary}{Corollary}
\newtheorem{Claim}{Claim}
\newtheorem{Remark}{Remark}
\def\e{\mathbb{ E}}
\usepackage{color}

\def\bit{\bibitem}

\def\p{\partial}
\def\d{\mathrm{d}}

\usepackage{extarrows}
\def\eqt#1{\xlongequal{\!\!#1\!\!}} % equal sign with text above

\let\ifexpand\iffalse

\begin{document}

\title{Higher Order Derivatives in Costa's Entropy Power Inequality }

\author{Fan Cheng,~\IEEEmembership{Member,~IEEE}
       and Yanlin Geng,~\IEEEmembership{Member,~IEEE}

\thanks{F.\ Cheng  was with the Institute of Network Coding, The Chinese University of Hong Kong, N.T., Hong Kong. He is now with the department of ECE, NUS, Singapore. Email: chengfan85@gmail.com}

\thanks{Y.\ Geng  was with the Department of Information Engineering, The Chinese University of Hong Kong, N.T., Hong Kong.  Now he is with the School of Information Science and Technology, ShanghaiTech University, China. Email: gengyanlin@gmail.com}

\thanks{The work of F. Cheng  was partially funded by a grant from the University Grants Committee of the Hong Kong Special Administrative Region (Project No.\ AoE/E-02/08) and Shenzhen Key Laboratory of Network Coding Key Technology and Application, Shenzhen, China (ZSDY20120619151314964). 
{ The work of Y.  Geng was  partially supported by a GRF grant from the University Grants Committee of the Hong Kong Special Administrative Region (Project No. 2150743) and the Science and Technology Commission of Shanghai Municipality (15YF1407900). } 
This paper was presented in part at IEEE Iran Workshop on Communication and Information Theory, 2014~\cite{CGFisher}.}
}

\maketitle

\begin{abstract}
Let $X$ be an arbitrary continuous random variable and $Z$ be an independent Gaussian random variable with zero mean and unit variance. For $t~>~0$, Costa proved that $e^{2h(X+\sqrt{t}Z)}$ is concave in $t$, where the proof hinged on the first and second order derivatives of $h(X+\sqrt{t}Z)$. Specifically, these two derivatives are signed, i.e., $\frac{\p}{\p t}h(X+\sqrt{t}Z) \geq 0$ and $\frac{\p^2}{\p t^2}h(X+\sqrt{t}Z) \leq 0$.
In this paper, we show that  the third order derivative of $h(X+\sqrt{t}Z)$  is nonnegative, which implies that the Fisher information $J(X+\sqrt{t}Z)$ is convex in $t$.
We further show that the fourth order derivative of $h(X+\sqrt{t}Z)$  is nonpositive.   Following the first four derivatives, we make two conjectures on $h(X+\sqrt{t}Z)$: the first is that $\frac{\partial^n}{\partial t^n} h(X+\sqrt{t}Z)$ is nonnegative in $t$ if $n$ is odd, and nonpositive otherwise; the second is that $\log J(X+\sqrt{t}Z)$ is convex in $t$. The first conjecture can be rephrased in the context of completely monotone functions: $J(X+\sqrt{t}Z)$ is completely monotone in $t$.  The history of the first conjecture may date back to  a problem in mathematical physics studied by McKean in 1966.  Apart from these results, we  provide a geometrical interpretation to the covariance-preserving transformation and study the concavity of $h(\sqrt{t}X+\sqrt{1-t}Z)$, revealing its connection with Costa's EPI.  
%A conjecture on the convexity of $J(\sqrt{t}X +\sqrt{1-t}Z)$ is introduced corresponding to the convexity of   $\log J(X+\sqrt{t}Z)$.

%The convexity/concavity of information on the probability distribution $\sqrt{t} X + \sqrt{1-t}Z$ is also studied, showing that $h(\sqrt{t} X + \sqrt{1-t}Z)$ is concave in $t$ and it is equivalent to Costa's EPI. We %conjecture that $J(\sqrt{t} X + \sqrt{1-t}Z)$ is convex in $t$, which is highly related to the convexity of $\log J(X+\sqrt{t}Z)$.
\end{abstract}


%\doublecolumn

%\newpage

\section{Introduction}
%\footnote{In general, the integrals may not exist. In this work, since for $t>0$ and the Gaussian noise $Z$ the density  of $X+\sqrt{t}Z$ is smooth, all the integrals  exist for any $X$.}
\IEEEPARstart{F}{or}
a continuous random variable $X$ with density $g(x)$, the differential entropy is defined as
\begin{equation}
h(X) := -\int_{-\infty}^{+\infty} g(x)\log g(x) \d x, \label{diff-entropy}
\end{equation}
where $\log$ is the natural logarithm.
The Fisher information (e.g., Cover \cite[p. 671]{Cover-Ele}) is defined as
\begin{equation}
J(X) := \int_{-\infty}^{+\infty} g(x)\left[\frac{\frac{\partial}{\partial x} g(x) }{g(x)}\right]^2 \d x. \label{fisher-inform}
\end{equation}

The entropy power inequality (EPI) introduced by Shannon~\cite{Shannon48} states that for any two independent continuous random variables $X$ and $Y$,
\begin{equation}\label{ShannonEPI}
e^{2h(X+Y)} \geq e^{2h(X)} + e^{2h(Y)},
\end{equation}
where the equality holds if and only if both $X$ and $Y$ are Gaussian.

Shannon did not give a proof and there was a gap in his argument. The first rigorous proof was made by Stam in \cite{Stam59}, where he applied an equality that connected  Fisher information and  differential entropy and the so-called Fisher information inequality (FII) was proved; i.e.,
 \begin{equation}\notag
 \frac{1}{J(X+Y)}\geq \frac{1}{J(X)} + \frac{1}{J(Y)}.
 \end{equation}
Later, Stam's proof was simplified by Blachman~\cite{Blachman65}. Zamir~\cite{ZamirFII} proved the FII via a data processing argument in Fisher information.
Lieb \cite{Lieb78} showed an equivalent form of EPI and proved the equivalent form via Young's inequality. Lieb's argument has been widely used as a common step in the subsequent proofs of EPI. Recently, Verd\'{u} and Guo \cite{VG-EPI06} gave a proof by invoking an equality that related  minimum mean square error estimation and  differential entropy. Rioul \cite{Rioul11} devised a Markov chain on $X$, $Y$, and the additive Gaussian noise, from which EPI can be proved via the data processing inequality and properties of mutual information.

There are several generalizations of EPI. Costa \cite{CostaEPI} proved that the entropy power $e^{2h(X+\sqrt{t}Z)}$ is concave in $t$, where the first and second order derivatives of $h(X+\sqrt{t}Z)$ were obtained. Moreover, these two derivatives are signed, i.e., positive or negative.
Dembo \cite{Dembo89} gave a simple proof to Costa's EPI via FII. 
Villani \cite{VillaniEPI} simplified the proof in~\cite{CostaEPI} by using some advanced techniques as well as the heat equation noticed by \cite{Stam59}, which is  instrumental in our work. The generalization of EPI in matrix form was obtained in Zamir and Feder \cite{ZamirEPI}. Liu and Viswanath \cite{LiuViswanath-EPI} generalized EPI by considering a covariance-constrained optimization problem which was motivated by multi-terminal coding problems.
Wang and Madiman~\cite{WangMadiman11} discussed EPI from the perspective of rearrangement. 

As one of the most important information inequalities, EPI (FII) has numerous proofs, generalizations, and applications.  In Barron \cite{Barron86}, FII was employed to strengthen the central limit theorem. The relationships of EPI to  inequalities  in other branches of mathematics can be found in Dembo \textit{et al.}  \cite{DemboCover91}. The literature is so vast that instead of trying to be complete, we only mention the results that are most relevant to our discussion.   A recent comprehensive survey can be found in~\cite{Rioul11}, and the book by El~Gamal and Kim \cite{GamalKim-NIT} also serves as a very good repository. 

In this paper, inspired by \cite{CostaEPI},  we make some   progress  and introduce related conjectures which reveal  even more fundamental facts about Gaussian random variables in the view of information theory. By harnessing the power of the techniques in \cite{CostaEPI}, i.e, heat equation and integration by parts, we obtain the third and fourth order derivatives of $h(X+\sqrt{t}Z)$, which are also signed. Summarizing all the derivatives of  $h(X+\sqrt{t}Z)$, we conjecture that $\frac{\p^n}{\p t^n}h(X+\sqrt{t}Z)$ is signed for any $n$. Corresponding to Costa's EPI,  we further conjecture that $\log J(X+\sqrt{t}Z)$ is convex in $t$. We investigate the concavity of $h(\sqrt{t}X+\sqrt{1-t}Z)$, showing that it is concave in $t$ and is equivalent to Costa's EPI. We  provide a geometrical interpretation to the covariance-preserving transformation. The connection between the convexities of $J(\sqrt{t}X+\sqrt{1-t}Z)$ and $\log J(X+\sqrt{t}Z)$
is also revealed. Finally, we state some results from the literature, including McKean's problem and completely monotone functions.

The paper is organized as follows. In Section~\ref{sec:1}, we introduce the background and the main result on derivatives. In Section~\ref{sec:2}, some preliminaries are stated. In Section~\ref{sec--10} and \ref{sec:proof-deri-4}, the derivatives are verified. We discuss the uniqueness of the signed form in Section~\ref{sec--13}. The conjecture is introduced in Section~\ref{sec--12}.  In Section~\ref{sec--14}, we give a geometrical interpretation to the covariance-preserving transformation and  prove an inequality which is equivalent to Costa's EPI. In Section~\ref{sec:FD}, we discuss some further issues. We conclude the paper in Section~\ref{sec:3}.

\section{The high-order derivatives}\label{sec:1}
Consider a random variable $X$ with density $g(x)$, and an independent standard Gaussian random variable $Z$, denoted as $Z\sim \mathcal{N}(0,1)$. For $t\geq 0$, let
\begin{equation}\notag
Y_t := X+\sqrt{t}Z.
\end{equation}
The density of $Y_t$ is
\begin{equation}\notag
f(y,t) = \int_{-\infty}^{+\infty}g(x)\frac{1}{\sqrt{2\pi t}}e^{-\frac{(y-x)^2}{2t}}\d x.
\end{equation}
\underline{Notation}: For the derivatives, in addition to the usages of  $f_{yy}$, $f_{yt}$ and $\frac{\p^2 }{\p y^2}f$, by $f^{(n)}$ we always mean
\begin{equation}\notag
f^{(n)} := \frac{\p^n}{\p y^n} f.
\end{equation}
Sometimes, {for ease of notation we also denote
\begin{align*}
f_n := f^{(n)} = \frac{\p^n}{\p y^n} f.
\end{align*}
}
The integration interval, usually $(-\infty,+\infty)$, will be omitted, unless it is not clear from the context.

In this paper, the main result is the following two theorems.
\begin{Theorem} \label{mainTheorem}
For  $t>0$,
\begin{align}
\frac{\p^3}{\p t^3}h(Y_t) %\nonumber \\
%&= \frac{1}{2}\int f\left( \frac{f_{yyy}}{f}-\frac{f_{y}f_{yy}}{f^2}+\frac{1}{3}\frac{f_y^3}{f^3}\right)^2  + \frac{f_{y}^4f_{yy}}{36f^4}   \d y \label{eq05}\\
&= \frac{1}{2}\int f\left( \frac{f_{3}}{f}-\frac{f_{1}f_{2}}{f^2}+\frac{1}{3}\frac{f_1^3}{f^3}\right)^2  +  \frac{f_1^6}{45f^5} \d y. \label{eq06666}
\end{align}
This implies that $J(Y_t)$ is convex in $t$.
\end{Theorem}

\begin{Theorem}\label{mainTheorem-h4}
For  $t>0$,
\begin{align}
&\frac{\p^4}{\p t^4}h(Y_t)\notag \\
&= -\frac{1}{2}\int f \left( \frac{f_4}{f} -\frac{6}{5} \frac{f_1f_3}{f^2} -\frac{7}{10} \frac{f_2^2}{f^2} +\frac{8}{5} \frac{f_1^2f_2}{f^3} -\frac{1}{2} \frac{f_1^4}{f^4} \right)^2 \notag \\
& \indent  + f \left(\frac{2}{5} \frac{f_1f_3}{f^2}  -\frac{1}{3} \frac{f_1^2f_2}{f^3} + \frac{9}{100} \frac{f_1^4}{f^4}\right)^2 \notag\\
& \indent  + f \left( -\frac{4}{100} \frac{f_1^2f_2}{f^3}  + \frac{4}{100} \frac{f_1^4}{f^4}\right)^2 \notag\\
& \indent  + \frac{1}{300}   \frac{f_2^4}{f^3} +\frac{56}{90000}  \frac{f_1^4f_2^2}{f^5} + \frac{13}{70000}  \frac{f_1^8}{f^7} \d y.
\end{align}
This implies that $\frac{\p^4}{\p t^4}h(Y_t)\leq 0$.
\end{Theorem}
Our work is highly related to the following theorem.
\begin{Theorem}[Costa's EPI \cite{CostaEPI}]\label{Costaepi}
$e^{2h(Y_t)}$ is concave in $t$, where $t>0$.
\end{Theorem}
There are several methods to prove Costa's EPI, and a straightforward way is to calculate the first and second order derivatives of $h(Y_t)$ and show some inequality holds.
The expressions on $\frac{\p }{\p t} h(Y_t)$ and $\frac{\p^2 }{\p t^2} h(Y_t)$ are already obtained in Lemma~\ref{lemma:1}.
\begin{Lemma}\label{lemma:1}
\begin{align}
& \frac{\p}{\p t} h(Y_t) = \frac{1}{2}J(Y_t); \label{lemma:1-1} %\quad \text{ (de Bruijn's identity) }
\\
& \frac{\p^2 }{\p t^2}h(Y_t) = -\frac{1}{2}\int f \left(\frac{f_{yy}}{f}-\frac{f_y^2}{f^2}\right)^2\d y. \label{lemma:1-2}
\end{align}
\end{Lemma}
The proof can be found in \cite{CostaEPI, VillaniEPI}. The first equation is called de Bruijn's identity in the literature and is due to de Bruijn.
Using Lemma~\ref{lemma:1}, one can readily show that $\frac{\p }{\p t} h(Y_t)$~$\geq$~$0$ and $\frac{\p^2 }{\p t^2} h(Y_t)\leq 0$.
In Theorem \ref{mainTheorem}, we have presented the expressions of $\frac{\p^3 }{\p t^3} h(Y_t)$ and showed that $\frac{\p^3 }{\p t^3} h(Y_t)\geq 0$. A much more complicated result on $\frac{\p^4 }{\p t^4} h(Y_t)$ is stated in Theorem~\ref{mainTheorem-h4}. We notice that in Guo \textit{et al.} \cite[Proposition 9]{GuoShamaiVerdu}, the third and fourth order derivatives of $h(\sqrt{t}X+Z)$ have been computed, but these derivatives cannot  determine the corresponding signs of $h(X+\sqrt t Z)$. However,  the signs of  $h(X+\sqrt t Z)$ are determined by Theorem~\ref{mainTheorem} and \ref{mainTheorem-h4}.

In the next section, we  introduce the necessary tools to prove Theorem~\ref{mainTheorem} and \ref{mainTheorem-h4}.

\section{Preliminaries}\label{sec:2}
The differential entropy and Fisher information may not be well defined due to the integration issue.  In the literature, there are no simple and general conditions which can guarantee their  existence (c.f. \cite{Rioul11}).
In general, the behavior of the differential entropy and Fisher information may be unpredictable as shown by Wu and Verd\'{u} \cite{WuVerduMMSE}.
However,  this work studies the higher order derivatives of $h(Y_t)$, where $t>0$ is imposed. Under this assumption, $Y_t$ has some good properties; e.g., in \cite{CostaEPI}, the density  of $Y_t$ is proved to be infinitely differentiable everywhere.

\subsection{Properties of $f(y,t)$}

The following property is well known (e.g., \cite[Lemma $1$]{Rioul11}). %[where?]           %[for example Rioul's paper]
\begin{Proposition}
For any fixed $t>0$ and any  $n\in \mathbb{Z}_+$, all the derivatives $f^{(n)}(y,t)$ exist, are bounded, and satisfy
\begin{align}
\lim_{|y|\to \infty} f^{(n)}(y,t) =0 .\notag
\end{align}
\end{Proposition}

The following property is used repeatedly in the rest of the paper, for dealing with integration by parts. The proof is presented in Appendix~\ref{apdx:prop-int-zero}.
\begin{Proposition}\label{prop-int-zero}
For any $r,m_i,k_i \in   \mathbb{Z}_{+}$, the following integral exists
\begin{align*}
\int f\left| \prod_{i=1}^{r} \frac{[f^{(m_i)}]^{k_i}}{f^{k_i}}\right| \d y.
\end{align*}
In particular, this implies that
{
\begin{align}\label{lim:term-zero}
\lim_{|y|\to \infty} f \prod_{i=1}^{r} \frac{[f^{(m_i)}]^{k_i}}{f^{k_i}}= 0.
\end{align}
}
\end{Proposition}

\subsection{The heat equation}
For a Gaussian random variable $\hat{X}\sim \mathcal{N}(\mu,\sigma^2)$ with density function $\hat{f}(x)$,
one can show that the following heat equation holds
$$
 \frac{\partial  }{\partial (\sigma^2)}\hat{f} = \frac{1}{2}\frac{\partial^2 }{\partial x^2}\hat{f}.
$$
The heat equation also holds for $Y_t$ \cite{Stam59}, and was used by \cite{VillaniEPI} to simplify Costa's proof.

\begin{Lemma}\label{lemma:heat}
\begin{equation}\label{heatEqn}
    \frac{\p }{\p t} f(y,t) = \frac{1}{2}\frac{\p^2 }{\p y^2} f(y,t).
\end{equation}
\end{Lemma}
\begin{IEEEproof}
{
The proof is known in the literature and we present it here for completeness.}
By some calculus,
\begin{align*}
 f_t &=  \int g(x)  \frac{1}{\sqrt{2\pi t}} e^{-\frac{(y-x)^2}{2t}}  \left(\frac{(y-x)^2}{2}\frac{1}{t^2}-\frac{1}{2 t}\right) \d x,  \\
 f_y     &=   \int g(x)\frac{1}{\sqrt{2\pi t}}e^{-\frac{(y-x)^2}{2t}}\left(-\frac{1}{t}(y-x)\right) \d x, \\
 f_{yy}  &= \int g(x)\frac{1}{\sqrt{2\pi t}}e^{-\frac{(y-x)^2}{2t}}\left[\left(\frac{1}{t}(y-x)\right)^2-\frac{1}{t}\right] \d x.
\end{align*}
By comparing $f_{yy}$ with $f_t$, the lemma can be proved.
\end{IEEEproof}

%\blue{
%The proofs of Lemma \ref{lemma:1}  and \ref{lemma:heat} are known in literature and are the origin of our work.  We defer them to the Appendix~\ref{proof-lemma1} and \ref{proof-lemma-heat}, respectively. In the next section, following Lemma \ref{lemma:1}  and \ref{lemma:heat}, we  derive the third order derivative of $h(Y_t)$; i.e., to prove Theorem \ref{mainTheorem}.  The main strategy to prove Theorem 1 and 2 is that we use heat equation to replace $t$ by $y$. Then we use integration by parts to simplify the integral items in the derivatives  until a signed form can be obtained.
%}

\subsection{Proof to Lemma~\ref{lemma:1}}
\label{sec:proof-lem-d1-d2}
{
The proof to Lemma~\ref{lemma:1} is known in the literature. Here we slightly modify the proof, so that the idea carries over to the proof of Theorem \ref{mainTheorem} and even the cases with  higher-order derivatives.
}
\begin{IEEEproof}
For the first order derivative we have
\begin{align*}
\frac{\partial }{\partial t} h(Y_t)
&= \frac{\p}{\p t}\left[-\int f(y,t)\log f(y,t)\d y\right] \\
&=  -\int f_t(1 + \log f) \d y \\
&\eqt{\eqref{heatEqn}} - \int \frac12 f_{yy} (1+ \log f) \d y \notag\\ % \quad \text{ (by \eqref{heatEqn}) }
&= -\frac12 \int (1+\log f) \d f_y\notag\\
&\eqt{(a)} -\frac12 f_y (1+ \log f) \bigg |_{y=-\infty}^{+\infty}  + \frac12 \int \frac{f_y^2}{f} \d y  \\
&\eqt{(b)}  0+ \frac{1}{2} \int \frac{f_y^2}{f}\d y \\
&=  \frac{1}{2}J(Y_t).
\end{align*}
In $(a)$ we apply integration by parts. In $(b)$ the limits are zero, because $f_{y}(1+\log f)= \frac{f_{y}}{\sqrt{f}} (\sqrt{f}+\sqrt{f}\log f)$, where $\frac{f_y^2}{f}\to 0$ from Proposition~\ref{prop-int-zero}, $\sqrt{f}\to 0$ as $|y|\to \infty$, and $\sqrt{f}\log f\to 0$ because $x\log x \to 0$ as $x\to 0$.

For the second order derivative, similarly
\begin{align*}
2 \frac{\p^2 }{\p t^2}h(Y_t)
&= \int \frac{2f_y f_{yt}f - f^2_yf_t}{f^2} \d y \\
&\eqt{\eqref{heatEqn}} \int \frac{f_y f_{yyy}}{f}  - \frac{f^2_yf_{yy}}{2f^2} \d y.
\end{align*}
For the second term
\begin{align*}
\int \frac{f_y^2f_{yy}}{f^2} \d y
&= \int \frac{f_y^2}{f^2} \d f_y \notag\\
&= \frac{f_y^3}{f^2}\bigg |_{y=-\infty}^{+\infty} -\int f_y  2\frac{f_y}{f} \frac{f_{yy}f-f_yf_y}{f^2} \d y \\
&\eqt{\eqref{lim:term-zero}} 0  -2\int \frac{f_y^2f_{yy}}{f^2}\d y + 2\int \frac{f_y^4}{f^3} \d y.
\end{align*}
Hence
\begin{align}
\int \frac{f_y^2 f_{yy}}{f^2}\d y =\int \frac{2f_{y}^{4}}{3f^3} \d y. \label{eqn:1221}
\end{align}
For the first term
\begin{align}
\int \frac{f_y f_{yyy}}{f} \d y
&= \int \frac{f_y}{f} \d f_{yy} \notag \\
&= \frac{f_yf_{yy}}{f} \bigg|_{y=-\infty}^{+\infty} - \int f_{yy} \frac{f_{yy}f-f_yf_y}{f^2} \d y \notag \\
&\eqt{\eqref{lim:term-zero}} 0 +\int -\frac{f_{yy}^2}{f} +\frac{f_y^2f_{yy}}{f^2} \d y \notag \\
&\eqt{\eqref{eqn:1221}} \int -\frac{f_{yy}^2}{f} +\frac{2f_{y}^{4}}{3f^3} \d y. \label{eqn:1311}
\end{align}
Combining these two terms we have
\begin{align}
2 \frac{\p^2 }{\p t^2}h(Y_t)
&= \int \frac{f_yf_{yyy}}{f} - \frac{f^2_yf_{yy}}{2f^2} \d y \notag \\
&\eqt{\eqref{eqn:1311}\eqref{eqn:1221}} \int -\frac{f_{yy}^2}{f} +\frac{f_{y}^{4}}{3f^3} \d y. \label{eqn:deri-2-final}
\end{align}

Now it suffices to show that the right-hand side term in \eqref{lemma:1-2} has the same form:
\begin{align*}
 -\int f \left(\frac{f_{yy}}{f}-\frac{f_y^2}{f^2}\right)^2\d y
&= \int -\frac{f_{yy}^2}{f} +\frac{2f_y^2f_{yy}}{f^2} -\frac{f_y^4}{f^3} \d y \\
&\eqt{\eqref{eqn:1221}} \int -\frac{f_{yy}^2}{f} +\frac{4f_y^4}{3f^3} -\frac{f_y^4}{f^3} \d y \\
&= \eqref{eqn:deri-2-final}.
\end{align*}
Thus the proof is finished.
\end{IEEEproof}

One may notice that we first use the heat equation to deal with $f_t$, then apply integration by parts to eliminate those terms whose highest-order derivatives have power one. Equation \eqref{eqn:1311} explains this elimination, as one can see that in the final expression the highest-order derivatives are $f_{yy}^2$ and $f_y^4$, whose powers are bigger than one.

%%%%%%%%%%%%%%%%%%%%%%%%%%%%%%%%%%%%%%%%%%%%%%%%%%%%%%%%%%%%%%%%%%%%%
\section{Proof to Theorem~\ref{mainTheorem}} \label{sec--10}

The following lemma is instrumental in proving Theorem~\ref{mainTheorem}.
\begin{Lemma}\label{lem:deri-3}
\begin{align}
\int \frac{f_1^4f_2}{f^4} \d y &= \int \frac{4f_1^6}{5f^5} \d y \label{eqn:1241} \\
\int \frac{f_1^3f_3}{f^3} \d y %&= \int - \frac{3f_1^2f_2^2}{f^3} +\frac{3f_1^4f_2}{f^4} \d y \notag \\
&= \int - \frac{3f_1^2f_2^2}{f^3} +\frac{12f_1^6}{5f^5} \d y \label{eqn:1331} \\
\int \frac{f_1f_2f_3}{f^2} \d y &= \int -\frac{f_2^3}{2f^2} + \frac{f_1^2f_2^2}{f^3} \d y \label{eqn:123111} \\
\int \frac{f_2f_4}{f}  \d y  %&= \int -\frac{f_3^2}{f} + \frac{f_1f_2f_3}{f^2} \d y  \notag \\
&= \int -\frac{f_3^2}{f} -\frac{f_2^3}{2f^2} + \frac{f_1^2f_2^2}{f^3} \d y \label{eqn:2411}
\end{align}
\end{Lemma}
\begin{IEEEproof}
See Appendix~\ref{app:lem-deri-3}.
\end{IEEEproof}

This lemma is similar to what we did in equations \eqref{eqn:1221} and \eqref{eqn:1311}: For the terms on the left-hand side, the highest-order derivatives have power one; while for the right-hand side, they are bigger than one.

Next, we prove Theorem~\ref{mainTheorem}.
\begin{IEEEproof}
From  \eqref{eqn:deri-2-final}
\begin{align*}
2\frac{\p^2 h(Y_t)}{\p t^2} &= \int - \frac{f_{yy}^2}{f} +  \frac{f_{y}^{4}}{3f^3} \d y
\equiv \int - \frac{f_{2}^2}{f} +  \frac{f_{1}^{4}}{3f^3} \d y.
\end{align*}
Thus
\begin{align*}
2\frac{\p^3 h(Y_t)}{\p t^3}
&= \int \left( - \frac{f_{2}^2}{f} +  \frac{f_{1}^{4}}{3f^3} \right)_t\d y.
\end{align*}
By repeatedly applying the heat equation,
\begin{align*}
 \int \left(\frac{f_{2}^2}{f}\right)_t  \d y
&= \int \frac{2f_{2}f_{2t}f-f_{2}^2f_t}{f^2}\d y \\
&\eqt{\eqref{heatEqn}} \int \frac{2f_{2}\frac12 f_{4}f -f_{2}^2 \frac12 f_{2}}{f^2}\d y \\
&= \int \frac{f_2f_4}{f} - \frac{f_2^3}{2f^2} \d y
\end{align*}

\begin{align*}
\int  \left(\frac{f_{1}^4}{3f^3}\right)_t  \d y
&=  \int  \frac{4f_{1}^3f_{1t}f^3-f_{1}^4 3f^2f_t}{3f^6}\d y \\
&\eqt{\eqref{heatEqn}}  \int  \frac{4f_{1}^3\frac12 f_{3}f^3-f_{1}^4 3f^2 \frac12 f_{2}}{3f^6}\d y \\
&=  \int  \frac{2f_{1}^3f_{3}}{3f^3} - \frac{f_{1}^4f_{2}}{2f^4} \d y.
\end{align*}
Substitute these terms and use Lemma~\ref{lem:deri-3}:
\begin{align}
 2\frac{\p^3h(Y_t)}{\p t^3}
&= \int \left( -\frac{f_2f_4}{f} + \frac{f_2^3}{2f^2} \right) +\left( \frac{2f_{1}^3f_{3}}{3f^3} - \frac{f_{1}^4f_{2}}{2f^4} \right) \d y \notag \\
& \eqt{\text{Lemma}~\ref{lem:deri-3}}
\int  -\left( -\frac{f_3^2}{f} -\frac{f_2^3}{2f^2} + \frac{f_1^2f_2^2}{f^3} \right) + \frac{f_2^3}{2f^2} \notag \\
&\quad +\frac23 \left( - \frac{3f_1^2f_2^2}{f^3} +\frac{12f_1^6}{5f^5} \right)  -\frac12 \left(\frac{4f_1^6}{5f^5}\right) \d y \notag \\
&= \int  \frac{f_3^2}{f} +\frac{f_2^3}{f^2} -\frac{3f_1^2f_2^2}{f^3}  +\frac{6f_1^6}{5f^5} \d y
\label{eqn:deri-3-final}
\end{align}

Then we do the same manipulations to $2\frac{\p^3}{\p t^3}h(Y_t)$ in Theorem~\ref{mainTheorem}. That is, applying Lemma~\ref{lem:deri-3} to the corresponding terms and we have
\begin{align*}
& \int f\left( \frac{f_3}{f}-\frac{f_1f_2}{f^2}+\frac{1}{3}\frac{f_1^3}{f^3}\right)^2  +  \frac{f_1^6}{45f^5} \d y \\
&= \int \frac{f_3^2}{f}+  \frac{f_1^2f_2^2}{f^3}+ \frac{f_1^6}{9f^5} -  \frac{2f_1f_2f_3}{f^2} \\
&\quad  + \frac{2f_1^3f_3}{3f^3} - \frac{2f_1^4f_2}{3f^4}  + \frac{f_1^6}{45f^5} \d y \\
& \eqt{\text{Lemma}~\ref{lem:deri-3}}
\int \frac{f_3^2}{f}+  \frac{f_1^2f_2^2}{f^3} + \frac{6f_1^6}{45f^5} -  2\left( -\frac{f_2^3}{2f^2} + \frac{f_1^2f_2^2}{f^3} \right) \\
&\quad   + \frac23\left(- \frac{3f_1^2f_2^2}{f^3} +\frac{12f_1^6}{5f^5}\right)- \frac23 \left(\frac{4f_1^6}{5f^5}\right) \d y \\
&= \int \frac{f_3^2}{f}-3  \frac{f_1^2f_2^2}{f^3} + \frac{54}{45}\frac{f_1^6}{f^5} +\frac{f_2^3}{f^2} \d y \\
&=\eqref{eqn:deri-3-final}
\end{align*}
Thus the expression is proved.

Finally,
\begin{equation}
\frac{\p^2}{\p t^2}J(Y_t) = 2 \frac{\p^3}{\p t^3} h(Y_t) \geq 0, \notag
\end{equation}
which means $J(Y_t)$ is convex in $t$.
\end{IEEEproof}

\section{Proof to Theorem~\ref{mainTheorem-h4}} \label{sec:proof-deri-4}
The proof is the same as that to Theorem~\ref{mainTheorem}, except there are more manipulations.
The following lemma is instrumental in proving Theorem~\ref{mainTheorem-h4}.
\begin{Lemma}\label{lem:deri-4}
\begin{align}
\int \frac{f_1^6f_2}{f^6} \d y &= \int \frac{6f_1^8}{7f^7} \d y \label{eqn:1261} \\
\int \frac{f_1^5f_3}{f^5} \d y &=  \int -\frac{5f_1^4f_2^2}{f^5} + \frac{30f_1^8}{7f^7} \d y \label{eqn:1351} \\
\int \frac{f_1^3f_2f_3}{f^4} \d y &= \int -\frac{3f_1^2f_2^3}{2f^4} + \frac{2f_1^4f_2^2}{f^5} \d y \label{eqn:123311} \\
\int \frac{f_1f_2^2f_3}{f^3} \d y &= \int -\frac{f_2^4}{3f^3} +\frac{f_1^2f_2^3}{f^4} \d y \label{eqn:123121} \\
\int \frac{f_1^4f_4}{f^4} \d y &= \int \frac{6f_1^2f_2^3}{f^4} - \frac{28f_1^4f_2^2}{f^5}  + \frac{120f_1^8}{7f^7} \d y\label{eqn:1441} \\
\int \frac{f_1^2f_2f_4}{f^3} \d y &= \int \frac{2f_2^4}{3f^3} - \frac{13f_1^2f_2^3}{2f^4} \notag \\
&\quad -\frac{f_1^2f_3^2}{f^3} + \frac{6f_1^4f_2^2}{f^5} \d y \label{eqn:124211} \\
\int \frac{f_2^2f_4}{f^2} \d y &= \int -\frac{2f_2f_3^2}{f^2} -\frac{2f_2^4}{3f^3} +\frac{2f_1^2f_2^3}{f^4} \d y  \label{eqn:2421} \\
\int \frac{f_1f_3f_4}{f^2} \d y &=  \int -\frac{f_2f_3^2}{2f^2} + \frac{f_1^2f_3^2}{f^3} \d y \label{eqn:134111} \\
\int \frac{f_3f_5}{f} \d y &= \int -\frac{f_4^2}{f} -\frac{f_2f_3^2}{2f^2} + \frac{f_1^2f_3^2}{f^3} \d y \label{eqn:3511}
\end{align}
\end{Lemma}
\begin{IEEEproof}
See Appendix~\ref{app:lem-deri-4}.
\end{IEEEproof}

Next, we prove Theorem~\ref{mainTheorem-h4}.
\begin{IEEEproof}
According to \eqref{eqn:deri-3-final}
\begin{align*}
 2\frac{\p^4 h(Y_t)}{\p t^4}
&= \int \left(\frac{f_3^2}{f} +\frac{f_2^3}{f^2}  -\frac{3f_1^2f_2^2}{f^3} +\frac{6f_1^6}{5f^5} \right)_t \d y
\end{align*}
We first apply the heat equation:
\begin{align*}
\int \left(\frac{f_3^2}{f}\right)_t \d y
&= \int \frac{2f_3f_{3t}f -f_3^2f_t}{f^2} \d y \\
&\eqt{\eqref{heatEqn}} \int \frac{f_3f_5f -f_3^2\frac12 f_2}{f^2} \d y \\
&= \int \frac{f_3f_5}{f} -\frac{f_2f_3^2}{2f^2} \d y
\end{align*}
\begin{align*}
\int \left(\frac{f_2^3}{f^2}\right)_t \d y
&= \int \frac{3f_2^2f_{2t}f^2 -f_2^32ff_t}{f^4} \d y  \\
&\eqt{\eqref{heatEqn}} \int \frac{\frac32 f_2^2f_4f^2 -f_2^3ff_2}{f^4} \d y \\
&= \int \frac{3f_2^2f_4}{2f^2} -\frac{f_2^4}{f^3} \d y
\end{align*}
\begin{align*}
& \int \left(\frac{3f_1^2f_2^2}{f^3}\right)_t \d y \\
&= \int \frac{6f_1f_2 (f_{1t}f_2+f_1f_{2t}) f^3 -3f_1^2f_2^2 3f^2f_t}{f^6} \d y  \\
&\eqt{\eqref{heatEqn}} \int \frac{3f_1f_2 (f_3f_2+f_1f_4) f^3 -\frac92 f_1^2f_2^2 f^2f_2}{f^6} \d y  \\
&= \int \frac{3f_1f_2^2f_3}{f^3} +\frac{3f_1^2f_2f_4}{f^3} -\frac{9f_1^2f_2^3}{2f^4} \d y
\end{align*}
\begin{align*}
 \int \left(\frac{6f_1^6}{5f^5} \right)_t \d y
&= \int \frac{36f_1^5f_{1t}f^5 - 6f_1^6 5f^4f_t}{5f^{10}} \d y  \\
&\eqt{\eqref{heatEqn}} \int \frac{ 18f_1^5f_3f^5 - 15 f_1^6 f^4f_2}{5f^{10}} \d y  \\
&= \int \frac{18f_1^5f_3}{5f^5} -\frac{3f_1^6f_2}{f^6} \d y
\end{align*}
Substitute these terms and use Lemma~\ref{lem:deri-4}:
\begin{align}
&2\frac{\p^4 h(Y_t)}{\p t^4} \notag \\
&= \int \left(  \frac{f_3f_5}{f} -\frac{f_2f_3^2}{2f^2} \right)
 +\left( \frac{3f_2^2f_4}{2f^2} -\frac{f_2^4}{f^3} \right) \notag \\
&\quad -\left( \frac{3f_1f_2^2f_3}{f^3} +\frac{3f_1^2f_2f_4}{f^3} -\frac{9f_1^2f_2^3}{2f^4} \right) \notag \\
&\quad +\left( \frac{18f_1^5f_3}{5f^5} -\frac{3f_1^6f_2}{f^6} \right) \d y \notag \\
&\eqt{\text{Lemma}~\ref{lem:deri-4}}
\int \left( -\frac{f_4^2}{f} -\frac{f_2f_3^2}{2f^2} + \frac{f_1^2f_3^2}{f^3} \right) -\frac{f_2f_3^2}{2f^2} \notag \\
&\quad +\frac32 \left( -\frac{2f_2f_3^2}{f^2} -\frac{2f_2^4}{3f^3} +\frac{2f_1^2f_2^3}{f^4} \right) -\frac{f_2^4}{f^3} \notag \\
&\quad -3 \left( -\frac{f_2^4}{3f^3} +\frac{f_1^2f_2^3}{f^4} \right) \notag \\
&\quad -3 \left( \frac{2f_2^4}{3f^3} - \frac{13f_1^2f_2^3}{2f^4}  -\frac{f_1^2f_3^2}{f^3} + \frac{6f_1^4f_2^2}{f^5}  \right) \notag \\
&\quad +\frac{9f_1^2f_2^3}{2f^4} +\frac{18}{5} \left( -\frac{5f_1^4f_2^2}{f^5} + \frac{30f_1^8}{7f^7} \right)
-3 \left(\frac{6f_1^8}{7f^7}\right) \d y \notag \\
&= \int -\frac{f_4^2}{f} +(-\frac12 -\frac12 -3) \frac{f_2f_3^2}{f^2} + (1+3) \frac{f_1^2f_3^2}{f^3} \notag \\
&\quad +(-1-1+1-2)\frac{f_2^4}{f^3} + (3-3+\frac{39}{2} +\frac92) \frac{f_1^2f_2^3}{f^4}  \notag \\
&\quad  +(-18-18)\frac{f_1^4f_2^2}{f^5} + (\frac{108}{7}-\frac{18}{7}) \frac{f_1^8}{f^7} \d y \notag \\
&= \int -\frac{f_4^2}{f}  -\frac{4f_2f_3^2}{f^2} + \frac{4f_1^2f_3^2}{f^3} -\frac{3f_2^4}{f^3} +\frac{24f_1^2f_2^3}{f^4}  \notag \\
&\quad  -\frac{36f_1^4f_2^2}{f^5} + \frac{90f_1^8}{7f^7} \d y
\label{eqn:deri-4-final}
\end{align}

Then we do the same manipulations to $2\frac{\p^4}{\p t^4}h(Y_t)$ in Theorem~\ref{mainTheorem-h4}. That is, applying Lemma~\ref{lem:deri-4} to the corresponding terms. To simplify the calculation, we first consider the following general expression
\begin{align}
& \int f \left(x_0 \frac{f_4}{f} + x_1 \frac{f_1f_3}{f^2} + x_2\frac{f_2^2}{f^2} + x_3\frac{f_1^2f_2}{f^3} + x_4\frac{f_1^4}{f^4}\right)^2 \d y \notag \\
& =  \int x_0^2 \frac{f_4^2}{f} + x_1^2\frac{f_1^2f_3^2}{f^3} + x_2^2\frac{f_2^4}{f^3} + x_3^2\frac{f_1^4f_2^2}{f^5}  + x_4^2\frac{f_1^8}{f^7} \notag \\
&\indent + 2x_0x_1\frac{f_1f_3f_4}{f^2} + 2x_0x_2\frac{f_2^2f_4}{f^2} + 2x_0x_3\frac{f_1^2f_2f_4}{f^3} \notag \\
&\indent  + 2x_0x_4\frac{f_1^4f_4}{f^4} + 2x_1x_2\frac{f_1f_2^2f_3}{f^3} + 2x_1x_3\frac{f_1^3f_2f_3}{f^4}  \notag\\
&\indent + 2x_1x_4\frac{f_1^5f_3}{f^5} + 2x_2x_3\frac{f_1^2f_2^3}{f^4} \notag\\
&\indent + 2x_2x_4\frac{f_1^4f_2^2}{f^5} + 2x_3x_4\frac{f_1^6f_2}{f^6}\d y \notag \\
& \eqt{\text{Lemma}~\ref{lem:deri-4}}
\int x_0^2 \frac{f_4^2}{f} + x_1^2\frac{f_1^2f_3^2}{f^3} + x_2^2\frac{f_2^4}{f^3} + x_3^2\frac{f_1^4f_2^2}{f^5} \notag\\
& \indent  + x_4^2\frac{f_1^8}{f^7}+ 2x_0x_1\left(-\frac{f_2f_3^2}{2f^2} + \frac{f_1^2f_3^2}{f^3}\right)\notag\\
&\indent + 2x_0x_2\left(-\frac{2f_2f_3^2}{f^2} -\frac{2f_2^4}{3f^3} +\frac{2f_1^2f_2^3}{f^4}\right) \notag\\
& \indent + 2x_0x_3\left(\frac{2f_2^4}{3f^3} - \frac{13f_1^2f_2^3}{2f^4} -\frac{f_1^2f_3^2}{f^3} + \frac{6f_1^4f_2^2}{f^5}\right)\notag \\
&\indent + 2x_0x_4\left(\frac{6f_1^2f_2^3}{f^4} - \frac{28f_1^4f_2^2}{f^5}  + \frac{120f_1^8}{7f^7}\right) \notag\\
& \indent + 2x_1x_2\left(-\frac{f_2^4}{3f^3} +\frac{f_1^2f_2^3}{f^4}\right) \notag\\
&\indent + 2x_1x_3\left(-\frac{3f_1^2f_2^3}{2f^4} + \frac{2f_1^4f_2^2}{f^5}\right) \notag\\
&\indent + 2x_1x_4\left(-\frac{5f_1^4f_2^2}{f^5} + \frac{30f_1^8}{7f^7} \right) \notag\\
&\indent + 2x_2x_3\frac{f_1^2f_2^3}{f^4} + 2x_2x_4\frac{f_1^4f_2^2}{f^5} + 2x_3x_4\left(\frac{6f_1^8}{7f^7}\right)\d y \notag \\
& =  \int x_0^2 \frac{f_4^2}{f} + (x_1^2+2x_0x_1-2x_0x_3)\frac{f_1^2f_3^2}{f^3} \notag\\
&\indent + (x_2^2-\frac{4}{3}x_0x_2+\frac{4}{3}x_0x_3-\frac{2}{3}x_1x_2) \frac{f_2^4}{f^3}\notag\\
&\indent + (x_3^2+12x_0x_3 - 56x_0x_4 + 4x_1x_3 \notag \\
&\qquad -10x_1x_4 +2x_2x_4) \frac{f_1^4f_2^2}{f^5}\notag \\
&\indent + (x_4^2+\frac{240}{7}x_0x_4 +\frac{60}{7}x_1x_4 +\frac{12}{7} x_3x_4)\frac{f_1^8}{f^7} \notag\\
&\indent + (-x_0x_1-4x_0x_2)\frac{f_2f_3^2}{f^2} \notag\\
&\indent + (4x_0x_2-13x_0x_3+12x_0x_4+2x_1x_2 \notag\\
&\qquad -3x_1x_3+2x_2x_3)\frac{f_1^2f_2^3}{f^4}\d y, \label{eqn:deri-4-general}
\end{align}
With this general simplification, we have
\begin{align}
  & \int f \left( \frac{f_4}{f} -\frac{6}{5} \frac{f_1f_3}{f^2} -\frac{7}{10} \frac{f_2^2}{f^2} +\frac{8}{5} \frac{f_1^2f_2}{f^3} -\frac{1}{2} \frac{f_1^4}{f^4} \right)^2 \d y \notag\\
& =     \int \frac{f_4^2}{f} + \left((-\frac{6}{5})^2+2(-\frac{6}{5})-2(\frac{8}{5})\right)\frac{f_1^2f_3^2}{f^3}\notag \\
& \indent + \bigg((-\frac{7}{10})^2-\frac{4}{3}(-\frac{7}{10})+\frac{4}{3}(\frac{8}{5}) \notag\\
& \indent  -\frac{2}{3}(-\frac{6}{5})(-\frac{7}{10})\bigg) \frac{f_2^4}{f^3} \notag\\
& \indent + \bigg ((\frac{8}{5})^2+12(\frac{8}{5}) - 56(-\frac{1}{2}) + 4(-\frac{6}{5})(\frac{8}{5}) \notag\\
& \indent  -10(-\frac{6}{5})(-\frac{1}{2}) +2(-\frac{7}{10})(-\frac{1}{2})\bigg)\frac{f_1^4f_2^2}{f^5}  \notag\\
& \indent + \bigg((-\frac{1}{2})^2+\frac{240}{7}(-\frac12)+\frac{60}{7}(-\frac65)(-\frac12) \notag\\
& \indent  +\frac{12}{7}(\frac85)(-\frac12) \bigg)\frac{f_1^8}{f^7} \notag\\
& \indent + \left(-(-\frac{6}{5})-4(-\frac{7}{10})\right)\frac{f_2f_3^2}{f^2} \notag \\
& \indent + \bigg(4(-\frac{7}{10})-13(\frac{8}{5})+12(-\frac{1}{2})+2(-\frac{6}{5})(-\frac{7}{10}) \notag\\
&\indent  -3(-\frac{6}{5})(\frac{8}{5})+2(-\frac{7}{10})(\frac{8}{5})\bigg)\frac{f_1^2f_2^3}{f^4}\d y \notag\\
& =\int \frac{f_4^2}{f} - \frac{104f_1^2f_3^2}{25f^3} + \frac{899f_2^4}{300f^3} + \frac{1839f_1^4f_2^2}{50f^5} \notag\\
&\indent -  \frac{1837f_1^8}{140f^7}+ \frac{4f_2f_3^2}{f^2} - \frac{122f_1^2f_2^3}{5f^4} \d y \label{eqn:deri-4-t5}
\end{align}

\begin{align}
  & \int f \left( \frac{2}{5} \frac{f_1f_3}{f^2} -\frac{1}{3} \frac{f_1^2f_2}{f^3}  + \frac{9}{100} \frac{f_1^4}{f^4}\right)^2\d y \notag \\
& =  \int \left((\frac{2}{5})^2\right)\frac{f_1^2f_3^2}{f^3} \notag \\
& \indent + \left((-\frac{1}{3})^2 + 4(\frac{2}{5})(-\frac{1}{3}) -10(\frac{2}{5})(\frac{9}{100}) \right)\frac{f_1^4f_2^2}{f^5} \notag \\
& \indent  + \left((\frac{9}{100})^2+\frac{60}{7}(\frac25)(\frac{9}{100}) +\frac{12}{7}(-\frac13)(\frac{9}{100}) \right)\frac{f_1^8}{f^7}  \notag\\
& \indent + \left(-3(\frac{2}{5})(-\frac{1}{3})\right)\frac{f_1^2f_2^3}{f^4}\d y \notag \\
& =  \int \frac{4f_1^2f_3^2}{25f^3} - \frac{704f_1^4f_2^2}{900f^5} + \frac{18567f_1^8}{70000f^7}  + \frac{2f_1^2f_2^3}{5f^4}\d y\label{eqn:deri-4-t3}
\end{align}

\begin{align}
& \int f \left( -\frac{4}{100} \frac{f_1^2f_2}{f^3}  + \frac{4}{100} \frac{f_1^4}{f^4}\right)^2\d y \notag\\
&  =     \int  \left((-\frac{4}{100})^2\right)\frac{f_1^4f_2^2}{f^5}\notag \\
& \indent + \left((\frac{4}{100})^2+\frac{12}{7}(-\frac{4}{100})(\frac{4}{100}) \right)\frac{f_1^8}{f^7}\d y \notag\\
& =  \int \frac{16f_1^4f_2^2}{10000f^5} - \frac{80f_1^8}{70000f^7}\d y \label{eqn:deri-4-t2}
\end{align}

By \eqref{eqn:deri-4-t5}, \eqref{eqn:deri-4-t3} and \eqref{eqn:deri-4-t2}
\begin{align}
& -\int f \left( \frac{f_4}{f} -\frac{6}{5} \frac{f_1f_3}{f^2} -\frac{7}{10} \frac{f_2^2}{f^2} +\frac{8}{5} \frac{f_1^2f_2}{f^3} -\frac{1}{2} \frac{f_1^4}{f^4} \right)^2  \notag\\
& \indent  + f \left(\frac{2}{5} \frac{f_1f_3}{f^2}  -\frac{1}{3} \frac{f_1^2f_2}{f^3} + \frac{9}{100} \frac{f_1^4}{f^4}\right)^2 \notag\\
& \indent  + f \left( -\frac{4}{100} \frac{f_1^2f_2}{f^3}  + \frac{4}{100} \frac{f_1^4}{f^4}\right)^2 \notag\\
& \indent  + \frac{1}{300}   \frac{f_2^4}{f^3} +\frac{56}{90000}  \frac{f_1^4f_2^2}{f^5} + \frac{13}{70000}  \frac{f_1^8}{f^7} \d y \notag\\
& =-\int \bigg(\frac{f_4^2}{f} - \frac{104f_1^2f_3^2}{25f^3} + \frac{899f_2^4}{300f^3} + \frac{1839f_1^4f_2^2}{50f^5}\notag \\
&\indent  -  \frac{1837f_1^8}{140f^7}+ \frac{4f_2f_3^2}{f^2} - \frac{122f_1^2f_2^3}{5f^4}\bigg) \notag \\
&\indent  +\left(\frac{4f_1^2f_3^2}{25f^3} - \frac{704f_1^4f_2^2}{900f^5} + \frac{18567f_1^8}{70000f^7}  + \frac{2f_1^2f_2^3}{5f^4}\right)\notag \\
&\indent  + \left(\frac{16f_1^4f_2^2}{10000f^5} - \frac{80f_1^8}{70000f^7}\right) \notag \\
&\indent  + \frac{1}{300}   \frac{f_2^4}{f^3} +\frac{56}{90000}  \frac{f_1^4f_2^2}{f^5} + \frac{13}{70000}  \frac{f_1^8}{f^7}\d y \notag\\
&= -\int \frac{f_4^2}{f}+ (-\frac{104}{25}+\frac{4}{25} )\frac{f_1^2f_3^2}{f^3} + (\frac{899}{300}+\frac{1}{300}) \frac{f_2^4}{f^3} \notag\\
&\indent  +(\frac{1839}{50}-\frac{704}{900}+\frac{16}{10000} +\frac{56}{90000} ) \frac{f_1^4f_2^2}{f^5} \notag\\
&\indent + (-\frac{1837}{140}+\frac{18567}{70000}-\frac{80}{70000} +  \frac{13}{70000} )\frac{f_1^8}{f^7} \notag \\
&\indent + \frac{4f_2f_3^2}{f^2} +(-\frac{122}{5}+\frac{2}{5}) \frac{f_1^2f_2^3}{f^4} \d y \notag \\
&= -\int \frac{f_4^2}{f}-4\frac{f_1^2f_3^2}{f^3} + 3 \frac{f_2^4}{f^3} +36 \frac{f_1^4f_2^2}{f^5} \notag\\
&\indent   -\frac{90}{7}\frac{f_1^8}{f^7}+ \frac{4f_2f_3^2}{f^2} -24 \frac{f_1^2f_2^3}{f^4} \d y \notag \\
&=\eqref{eqn:deri-4-final}, \notag
\end{align}
which completes the proof of Theorem~\ref{mainTheorem-h4}.
\end{IEEEproof}

%%%%%%%%%%%%%%%%%%

\ifexpand

\section{Proof to Theorem~\ref{mainTheorem-h4}} \label{sec--11}
By \eqref{eq06}, we obtain that
\begin{align}
& \frac{\p^4}{\p t^4} h(Y_t) \notag\\
& =\frac{1}{2}\int   \left(\frac{f_{3}^2}{f}\right)_t + \left(\frac{f_{1}^3f_{3}}{3f^3}\right)_t \notag \\
& \indent -\left(\frac{2f_1f_{2}f_{3}}{f^2}\right)_t   + \left(\frac{f_1^4f_{2}}{2f^4}\right)_t   \d y \label{4th-0-1}
\end{align}
In the following, we denote $\frac{\partial^n }{\partial y^n}f$ by $f_n$ for ease of notation. By repeatedly applying the heat equation to eliminate $f_t$,
\begin{align}
& \int \left(\frac{f_{3}^2}{f}\right)_t \d y \notag \\
& =\int \frac{(f_{3}^2)_tf-f_3^2f_t}{f^2} \d y \notag\\
& =\int \frac{2f_3f_{3t}f-f_3^2f_2/2}{f^2}\d y \notag\\
& =\int \frac{ff_3f_5-f_2f_3^2/2}{f^2}\d y \notag\\
& = \int \frac{f_3f_5}{f} -\frac{f_2f_3^2}{2f^2}\d y \label{4th-1}
\end{align}
Integration by parts to eliminate $f_5$:
\begin{align}
&\int \frac{f_3f_5}{f} dy \notag\\
&= \int \frac{f_3}{f}\d f_4 \notag\\
&= \frac{f_3f_4}{f}\bigg |_{-\infty}^{+\infty} -\int f_4 \left(\frac{f_3}{f}\right)_y \d y \\
&= -\int f_4\frac{f_4f-f_3f_1}{f^2} \d y \notag\\
&=\int -\frac{f_4^2}{f} + \frac{f_1f_3f_4}{f^2}  \d y \label{4th-2}
\end{align}
By \eqref{4th-1} and \eqref{4th-2},
\begin{align}
&\int \left(\frac{f_{3}^2}{f}\right)_t \d y \notag\\
&=  \int -\frac{f_4^2}{f} + \frac{f_1f_3f_4}{f^2}-\frac{f_2f_3^2}{2f^2} \d y \label{4th-2-1}
\end{align}
By repeatedly applying the heat equation,
\begin{align}
&  \int \left( \frac{f_{1}^3f_{3}}{3f^3} \right)_{t} \d y \notag\\
&= \int \frac{(f_{1}^3f_{3})_tf^3 - f_{1}^3f_{3} (f^3)_t}{3f^6} \d y \notag\\
&= \int \frac{((f_{1}^3)_tf_{3} + f_{1}^3f_{3t})f^3 - f_{1}^3f_{3}3f^2f_t}{3f^6} \d y \notag\\
&= \int  \frac{(3f_1^2f_{1t}f_{3} + f_{1}^3f_{3t})f^3 - f_{1}^3f_{3}3f^2f_t}{3f^6} \d y \notag\\
&= \int \frac{f_1^2f_3^2}{2f^3} + \frac{f_1^3f_5}{6f^3} - \frac{f_1^3f_2f_3}{2f^4} \d y \label{4th-3}
\end{align}
Integration by parts:
\begin{align}
& \int \frac{f_1^3f_5}{6f^3} \d y \notag\\
& = \int \frac{f_1^3}{6f^3} \d f_4 \notag\\
& =\frac{f_1^3f_4}{6f^3} \bigg |_{-\infty}^{+\infty} -\int \frac{f_4}{6} \left(\frac{f_1^3}{f^3}\right)_y \d y \\
& =  -\int \frac{f_4}{6}\times 3 \times \frac{f_1^2}{f^2}\frac{f_2f-f_1^2}{f^2} \d y \notag\\
& =  \int -\frac{f_1^2f_2f_4}{2f^3} + \frac{f_1^4f_4}{2f^4} \d y \label{4th-4}
\end{align}
By \eqref{4th-3} and \eqref{4th-4},
\begin{align}
&  \int \left( \frac{f_{1}^3f_{3}}{3f^3} \right)_{t} \d y \notag\\
& = \int \frac{f_1^2f_3^2}{2f^3} -\frac{f_1^2f_2f_4}{2f^3} + \frac{f_1^4f_4}{2f^4} - \frac{f_1^3f_2f_3}{2f^4} \d y \label{4th-4-1}
\end{align}
By repeatedly applying the heat equation,
\begin{align}
   &\int \left(-\frac{2f_1f_{2}f_{3}}{f^2}\right)_t  \d y \notag \\
&= \int -2 \frac{(f_1f_2f_3)_tf^2- f_1f_2f_3(f^2)_t}{f^4} \d y \notag\\
& =  \int -2 \frac{(f_{1t}f_2f_3+f_1f_{2t}f_3 + f_1f_2f_{3t})f^2- f_1f_2f_32ff_t}{f^4} \d y \notag\\
& =  \int -\frac{f_2f_3^2}{f^2}-\frac{f_1f_3f_4}{f^2}-\frac{f_1f_2f_5}{f^2} + \frac{2f_1f_2^2f_3}{f^3} \d y \label{4th-5}
\end{align}
Integration by parts:
\begin{align}
  & \int -\frac{f_1f_2f_5}{f^2} \d y \notag \\
& = \int -\frac{f_1f_2}{f^2} \d f_4  \notag \\
& = -\frac{f_1f_2f_4}{f^2}\bigg |_{-\infty}^{+\infty} +\int f_4 \left(\frac{f_1f_2}{f^2}\right)_y \d y\\
& =  \int f_4 \frac{(f_1f_2)_yf^2-f_1f_2(f^2)_y}{f^4} \d y  \notag \\
& =  \int f_4 \frac{(f_2^2+f_1f_3)f^2-f_1f_22ff_1}{f^4} \d y \notag \\
& =  \int \frac{f_2^2f_4}{f^2}  + \frac{f_1f_3f_4}{f^2} - \frac{2f_1^2f_2f_4}{f^3} \d y \label{4th-6}
\end{align}
By  \eqref{4th-5} and \eqref{4th-6},
\begin{align}
& \int \left(-\frac{2f_1f_{2}f_{3}}{f^2}\right)_t \d y \notag \\
& = \int -\frac{f_2f_3^2}{f^2} +\frac{f_2^2f_4}{f^2}   - \frac{2f_1^2f_2f_4}{f^3}+ \frac{2f_1f_2^2f_3}{f^3}\d y \label{4th-6-1}
\end{align}
By repeatedly applying the heat equation,
\begin{align}
  & \int \left(\frac{f_1^4f_{2}}{2f^4}\right)_t \d y \notag \\
& = \int \frac{(f_1^4f_2)_tf^4-f_1^4f_2(f^4)_t}{2f^8} \d y \notag  \\
& = \int \frac{((f_1^4)_tf_2+f_1^4f_{2t})f^4-f_1^4f_24f^3f_t}{2f^8} \d y \notag \\
& = \int \frac{(4f_1^3f_{1t}f_2+f_1^4f_{2t})f^4-f_1^4f_24f^3f_t}{2f^8} \d y \notag \\
& =\int  \frac{f_1^3f_2f_3}{f^4} + \frac{f_1^4f_4}{4f^4} -\frac{f_1^4f_2^2}{f^5} \d y \label{4th-6-2}
\end{align}

%% STOP 1
By \eqref{4th-0-1}, \eqref{4th-2-1}, \eqref{4th-4-1}, \eqref{4th-6-1} and \eqref{4th-6-2},
\begin{align}
&\frac{\p^4}{\p t^4} h(Y_t) \notag \\
&=\frac{1}{2}\int  \left(-\frac{f_4^2}{f} + \frac{f_1f_3f_4}{f^2}-\frac{f_2f_3^2}{2f^2}\right) \notag\\
&\indent +\left(\frac{f_1^2f_3^2}{2f^3} -\frac{f_1^2f_2f_4}{2f^3} + \frac{f_1^4f_4}{2f^4} - \frac{f_1^3f_2f_3}{2f^4}\right)\notag\\
&\indent +\left(-\frac{f_2f_3^2}{f^2} +\frac{f_2^2f_4}{f^2}   - \frac{2f_1^2f_2f_4}{f^3}+ \frac{2f_1f_2^2f_3}{f^3}\right) \notag\\
&\indent + \left(\frac{f_1^3f_2f_3}{f^4} + \frac{f_1^4f_4}{4f^4} -\frac{f_1^4f_2^2}{f^5}\right) \d y\notag\\
&= \frac{1}{2}\int   -\frac{f_4^2}{f} + \frac{f_1f_3f_4}{f^2}-\frac{3f_2f_3^2}{2f^2} + \frac{f_1^2f_3^2}{2f^3} -\frac{5f_1^2f_2f_4}{2f^3}\notag \\
&\indent + \frac{3f_1^4f_4}{4f^4} + \frac{f_1^3f_2f_3}{2f^4} +\frac{f_2^2f_4}{f^2} + \frac{2f_1f_2^2f_3}{f^3} -\frac{f_1^4f_2^2}{f^5}\d y \label{4thstop-1}
\end{align}
By \eqref{4thstop-1}, we have a general idea on the signed form of  $\frac{\p^4}{\p t^4} h(Y_t)$, which may have items like $\frac{f_4}{f}$, $\frac{f_1f_3}{f^2}$, $\frac{f_2^2}{f^2}$, $\frac{f_1^2f_2}{f^3}$, and $\frac{f_1^4}{f^4}$.
In what follows, we try to eliminate the higher order derivatives as many as possible via integration by parts.
\begin{align}
 & \int \frac{f_1f_3f_4}{f^2} \d y \notag \\
& = \int \frac{f_1f_3}{f^2} \d f_3 \notag\\
& = \int \frac{f_1}{2f^2} \d f_3^2 \notag\\
& = \frac{f_1f_3^2}{2f^2}\bigg |_{-\infty}^{+\infty}-\int \frac{f_3^2}{2}\left(\frac{f_1}{f^2}\right)_y \d y\\
& =  -\int \frac{f_3^2}{2}\frac{(f_1)_yf^2-f_1(f^2)_y}{f^4} \d y\notag\\
& = -\int \frac{f_3^2}{2} \frac{f_2f^2-f_12ff_1}{f^4}\d y \notag\\
& =  \int -\frac{f_2f_3^2}{2f^2} + \frac{f_1^2f_3^2}{f^3}\d y\label{f1f3f4}
\end{align}

%\begin{align}
%  &  \int \frac{f_1^2f_2f_4}{f^3} \d y = \int  \frac{f_1^2f_2}{f^3} \d f_3 = \frac{f_1^2f_2f_3}{f^3}\bigg |_{-\infty}^{+\infty} - \int f_3\left(\frac{f_1^2f_2}{f^3}\right)_y \d y\\
%= & \int  -f_3 \frac{(f_1^2f_2)_yf^3 - f_1^2f_2(f^3)_y}{f^6} = \int  -f_3 \frac{(2f_1f_2f_2+ f_1^2f_3)f^3 - f_1^2f_23f^2f_1}{f^6} \d y\\
%= & \int -\frac{2f_1f_2^2f_3}{f^3} - \frac{f_1^2f_3^2}{f^3} +\frac{3f_1^3f_2f_3}{f^4}  \d y
%\end{align}

\begin{align}
 &\int \frac{f_1f_2^2f_3}{f^3} \d y \notag\\
&= \int \frac{f_1f_2^2}{f^3} \d f_2  \notag\\
&= \int \frac{f_1}{3f^3}\d f_2^3 \notag\\
&= \frac{f_1f_2^3}{3f^3}\bigg |_{-\infty}^{+\infty} -\int \frac{f_2^3}{3}\left(\frac{f_1}{f^3}\right)_y \d y\\
&= - \int \frac{f_2^3}{3}\frac{f_2f^3-f_13f^2f_1}{f^6} \d y \notag\\
&= \int -\frac{f_2^4}{3f^3} +\frac{f_1^2f_2^3}{f^4}\d y \label{f1f22f3}
\end{align}

\begin{align}
& \int \frac{f_1^3f_2f_3}{f^4} \d y  \notag\\
& = \int \frac{f_1^3f_2}{f^4} \d f_2 \notag\\
& = \int \frac{f_1^3}{2f^4} \d f_2^2 \notag\\
& = \frac{f_1^3f_2^2}{2f^4}\bigg |_{-\infty}^{+\infty}   -\int \frac{f_2^2}{2}  \left(\frac{f_1^3}{f^4}\right)_y \d y \\
& = \int -\frac{f_2^2}{2}\frac{3f_1^2f_2f^4-f_1^34f^3f_1}{f^8} \d y\notag\\
& =\int -\frac{3f_1^2f_2^3}{2f^4} + \frac{2f_1^4f_2^2}{f^5}\d y \label{f13f2f3}
\end{align}

%\begin{align}
%\int - \frac{5f_1^2f_2}{2f^3} \d f_3 =
% \int \frac{5f_1f_2^2f_3}{f^3} + \frac{5f_1^2f_3^2}{2f^3} -\frac{15f_1^3f_2f_3}{2f^4}  \d y
%\end{align}

\begin{align}
&  \int \frac{f_1^2f_2f_4}{f^3} \d y \notag \\
& = \int  \frac{f_1^2f_2}{f^3} \d f_3 \notag\\
& = \frac{f_1^2f_2f_3}{f^3}\bigg |_{-\infty}^{+\infty} - \int f_3\left(\frac{f_1^2f_2}{f^3}\right)_y \d y \notag\\
& =  \int  -f_3 \frac{(f_1^2f_2)_yf^3 - f_1^2f_2(f^3)_y}{f^6} \notag\\
& = \int  -f_3 \frac{(2f_1f_2f_2+ f_1^2f_3)f^3 - f_1^2f_23f^2f_1}{f^6} \d y \notag\\
& =  \int -\frac{2f_1f_2^2f_3}{f^3} - \frac{f_1^2f_3^2}{f^3} +\frac{3f_1^3f_2f_3}{f^4}  \d y \notag\\
& =  \int -2\left(-\frac{f_2^4}{3f^3} +\frac{f_1^2f_2^3}{f^4}\right) - \frac{f_1^2f_3^2}{f^3} \notag \\ %\eqt{\eqref{f1f22f3}, \eqref{f13f2f3}}
& \indent \indent+ 3\left(-\frac{3f_1^2f_2^3}{2f^4} + \frac{2f_1^4f_2^2}{f^5}\right) \d y \label{f12f2f4-1}\\
& =  \int \frac{2f_2^4}{3f^3} - \frac{13f_1^2f_2^3}{2f^4} -\frac{f_1^2f_3^2}{f^3} + \frac{6f_1^4f_2^2}{f^5} \d y, \label{f12f2f4}
\end{align}
where \eqref{f12f2f4-1} is from \eqref{f1f22f3} and  \eqref{f13f2f3}.
%\begin{align}
%  & \int \frac{f_1^4f_4}{f^4} \d y =  \int \frac{f_1^4}{f^4} \d f_3 =\frac{f_1^4f_3}{f^4}\bigg |_{-\infty}^{+\infty} - \int  f_3 \left( \frac{f_1^4}{f^4} \right)_y \d y = \int - f_3 4\frac{f_1^3}{f^3}\frac{f_2f-f_1^2}{f^2}\d y\\
%= &\int -\frac{4f_1^3f_2f_3}{f^4} + \frac{4f_1^5f_3}{f^5}\d y
%\end{align}

%\begin{align}
% &\int \frac{f_2^2f_4}{f^2} \d y = \int \frac{f_2^2}{f^2} \d f_3 =  \frac{f_2^2f_3}{f^2}\bigg |_{-\infty}^{+\infty}  - \int  f_3 \left(\frac{f_2^2}{f^2}\right)_y \d y = \int - f_3 2\frac{f_2}{f}\frac{f_3f-f_2f_1}{f^2} \d y\\
%=& \int -\frac{2f_2f_3^2}{f^2} + \frac{2f_1f_2^2f_3}{f^3} \d y
%\end{align}

Integration by parts:
\begin{align}
& \int \frac{f_1^6f_2}{f^6}\d y \notag \\
& = \int \frac{f_1^6}{f^6}\d f_1  \notag \\
& =\int \frac{1}{7f^6}\d f_1^7  \notag \\
& =\frac{f_1^7}{7f^6} \bigg |_{-\infty}^{+\infty} - \int \frac{f_1^7}{7}\left(\frac{1}{f^6}\right)_y \d y \\
& = \int -\frac{f_1^7}{7}\frac{-6f_1}{f^7} \d y= \int \frac{6f_1^8}{7f^7} \d y \label{f16f2}
\end{align}

\begin{align}
& \int \frac{f_1^5f_3}{f^5} \d y\notag \\
& = \int \frac{f_1^5}{f^5}\d f_2 \notag\\
& = \frac{f_1^5f_2}{f^5}\bigg |_{-\infty}^{+\infty}  - \int f_2 \left(\frac{f_1^5}{f^5}\right)_y \d y  \\
& = \int -f_2 5\frac{f_1^4}{f^4}\frac{f_2f-f_1^2}{f^2} \d y\notag\\
& =  \int -\frac{5f_1^4f_2^2}{f^5} + \frac{5f_1^6f_2}{f^6} \d y \notag\\
& \eqt{\eqref{f16f2}} \int -\frac{5f_1^4f_2^2}{f^5} + \frac{30f_1^8}{7f^7} \d y \label{f15f3}
\end{align}

\begin{align}
& \int \frac{f_1^4f_4}{f^4} \d y \notag\\
& =  \int \frac{f_1^4}{f^4} \d f_3 \notag\\
& =\frac{f_1^4f_3}{f^4}\bigg |_{-\infty}^{+\infty} - \int  f_3 \left( \frac{f_1^4}{f^4} \right)_y \d y \notag\\
& = \int - f_3 4\frac{f_1^3}{f^3}\frac{f_2f-f_1^2}{f^2}\d y \notag\\
& = \int -\frac{4f_1^3f_2f_3}{f^4} + \frac{4f_1^5f_3}{f^5}\d y \notag\\
& =   \int -4\left( -\frac{3f_1^2f_2^3}{2f^4} + \frac{2f_1^4f_2^2}{f^5} \right) \notag\\
& \indent \indent + 4 \left( -\frac{5f_1^4f_2^2}{f^5} + \frac{30f_1^8}{7f^7}\right)\d y \label{f14f4-1}\\
 %= & \int \frac{6f_1^2f_2^3}{f^4} - \frac{28f_1^4f_2^2}{f^5}  + \frac{20f_1^6f_2}{f^6} \d y \\
& =  \int \frac{6f_1^2f_2^3}{f^4} - \frac{28f_1^4f_2^2}{f^5}  + \frac{120f_1^8}{7f^7} \d y, \label{f14f4}
\end{align}
where \eqref{f14f4-1} is from \eqref{f13f2f3} and \eqref{f15f3}.

Integration by parts:
\begin{align}
& \int \frac{f_2^2f_4}{f^2} \d y \notag \\
& = \int \frac{f_2^2}{f^2} \d f_3 \notag \\
& =\frac{f_2^2f_3}{f^2}  \bigg |_{-\infty}^{+\infty}  -  \int  f_3 \left(\frac{f_2^2}{f^2}\right)_y \d y \notag\\
& = \int - f_3 2\frac{f_2}{f}\frac{f_3f-f_2f_1}{f^2} \d y \notag\\
& = \int -\frac{2f_2f_3^2}{f^2} + \frac{2f_1f_2^2f_3}{f^3} \d y \notag\\
& \eqt{\eqref{f1f22f3}} \int -\frac{2f_2f_3^2}{f^2} + 2\left(-\frac{f_2^4}{3f^3} +\frac{f_1^2f_2^3}{f^4}\right) \d y \notag\\
& =  \int -\frac{2f_2f_3^2}{f^2} -\frac{2f_2^4}{3f^3} +\frac{2f_1^2f_2^3}{f^4} \d y \label{f22f4}
\end{align}

%%STOP 2:
By \eqref{4thstop-1}, \eqref{f1f3f4}, \eqref{f12f2f4}, \eqref{f14f4}, \eqref{f13f2f3}, \eqref{f22f4}, and \eqref{f1f22f3},
\begin{align}
& \frac{\p^4}{\p t^4} h(Y_t)  \notag\\
& = \frac{1}{2}\int   -\frac{f_4^2}{f} + \frac{f_1f_3f_4}{f^2}-\frac{3f_2f_3^2}{2f^2} + \frac{f_1^2f_3^2}{2f^3} \notag\\
& \indent -\frac{5f_1^2f_2f_4}{2f^3}   + \frac{3f_1^4f_4}{4f^4} + \frac{f_1^3f_2f_3}{2f^4} +\frac{f_2^2f_4}{f^2}  \notag\\
& \indent + \frac{2f_1f_2^2f_3}{f^3} -\frac{f_1^4f_2^2}{f^5}\d y \notag \\
& =  \frac{1}{2}\int-\frac{f_4^2}{f} + \left(-\frac{f_2f_3^2}{2f^2} + \frac{f_1^2f_3^2}{f^3}\right) \notag \\
&\indent -\frac{3f_2f_3^2}{2f^2} + \frac{f_1^2f_3^2}{2f^3} \notag\\
& \indent  -\frac{5}{2}\left(\frac{2f_2^4}{3f^3} - \frac{13f_1^2f_2^3}{2f^4} -\frac{f_1^2f_3^2}{f^3} + \frac{6f_1^4f_2^2}{f^5}\right) \notag\\
& \indent + \frac{3}{4}\left(\frac{6f_1^2f_2^3}{f^4} - \frac{28f_1^4f_2^2}{f^5}  + \frac{120f_1^8}{7f^7}\right)\notag \\
& \indent + \frac{1}{2}\left(-\frac{3f_1^2f_2^3}{2f^4} + \frac{2f_1^4f_2^2}{f^5}\right) \notag\\
& \indent +\left(-\frac{2f_2f_3^2}{f^2} -\frac{2f_2^4}{3f^3} +\frac{2f_1^2f_2^3}{f^4}\right) \notag\\
& \indent + 2\left(-\frac{f_2^4}{3f^3} +\frac{f_1^2f_2^3}{f^4}\right) -\frac{f_1^4f_2^2}{f^5}\d y \notag\\
& =\frac{1}{2}\int -\frac{f_4^2}{f}-\frac{4f_2f_3^2}{f^2} + \frac{4f_1^2f_3^2}{f^3} - \frac{3f_2^4}{f^3} \notag\\
&\indent +  \frac{24f_1^2f_2^3}{f^4} - \frac{36f_1^4f_2^2}{f^5} +  \frac{90f_1^8}{7f^7}  \d y \notag\\
& = - \frac{1}{2}\int \frac{f_4^2}{f}+\frac{4f_2f_3^2}{f^2} - \frac{4f_1^2f_3^2}{f^3} + \frac{3f_2^4}{f^3} \notag\\
&\indent   - \frac{24f_1^2f_2^3}{f^4} + \frac{36f_1^4f_2^2}{f^5} -  \frac{90f_1^8}{7f^7}  \d y \label{4thstop-2}
\end{align}
Now, we have greatly simplified the expression of $\frac{\p^4}{\p t^4} h(Y_t)$, which can help us to prove Theorem~\ref{mainTheorem-h4} subsequently.
First, we derive the following general expression for ease of calculation,
\begin{align}
& \int f \left(x_0 \frac{f_4}{f} + x_1 \frac{f_1f_3}{f^2} + x_2\frac{f_2^2}{f^2} + x_3\frac{f_1^2f_2}{f^3} + x_4\frac{f_1^4}{f^4}\right)^2 \d y \notag \\
& =  \int x_0^2 \frac{f_4^2}{f} + x_1^2\frac{f_1^2f_3^2}{f^3} + x_2^2\frac{f_2^4}{f^3} + x_3^2\frac{f_1^4f_2^2}{f^5} \notag \\
&\indent  + x_4^2\frac{f_1^8}{f^7} + 2x_0x_1\frac{f_1f_3f_4}{f^2} + 2x_0x_2\frac{f_2^2f_4}{f^2}  \notag \\
&\indent + 2x_0x_3\frac{f_1^2f_2f_4}{f^3} + 2x_0x_4\frac{f_1^4f_4}{f^4} + 2x_1x_2\frac{f_1f_2^2f_3}{f^3} \notag\\
&\indent + 2x_1x_3\frac{f_1^3f_2f_3}{f^4} + 2x_1x_4\frac{f_1^5f_3}{f^5} + 2x_2x_3\frac{f_1^2f_2^3}{f^4} \notag\\
&\indent + 2x_2x_4\frac{f_1^4f_2^2}{f^5} + 2x_3x_4\frac{f_1^6f_2}{f^6}\d y \notag \\
& =    \int x_0^2 \frac{f_4^2}{f} + x_1^2\frac{f_1^2f_3^2}{f^3} + x_2^2\frac{f_2^4}{f^3} \notag\\
& \indent + x_3^2\frac{f_1^4f_2^2}{f^5} + x_4^2\frac{f_1^8}{f^7}+ 2x_0x_1\left(-\frac{f_2f_3^2}{2f^2} + \frac{f_1^2f_3^2}{f^3}\right)\notag\\
&\indent + 2x_0x_2\left(-\frac{2f_2f_3^2}{f^2} -\frac{2f_2^4}{3f^3} +\frac{2f_1^2f_2^3}{f^4}\right) \notag\\
& \indent + 2x_0x_3\left(\frac{2f_2^4}{3f^3} - \frac{13f_1^2f_2^3}{2f^4} -\frac{f_1^2f_3^2}{f^3} + \frac{6f_1^4f_2^2}{f^5}\right)\notag \\
&\indent + 2x_0x_4\left(\frac{6f_1^2f_2^3}{f^4} - \frac{28f_1^4f_2^2}{f^5}  + \frac{120f_1^8}{7f^7}\right) \notag\\
& \indent + 2x_1x_2\left(-\frac{f_2^4}{3f^3} +\frac{f_1^2f_2^3}{f^4}\right) \notag\\
&\indent + 2x_1x_3\left(-\frac{3f_1^2f_2^3}{2f^4} + \frac{2f_1^4f_2^2}{f^5}\right) \notag\\
&\indent + 2x_1x_4\left(-\frac{5f_1^4f_2^2}{f^5} + \frac{30f_1^8}{7f^7} \right) \notag\\
&\indent + 2x_2x_3\frac{f_1^2f_2^3}{f^4} + 2x_2x_4\frac{f_1^4f_2^2}{f^5} + 2x_3x_4\left(\frac{6f_1^8}{7f^7}\right)\d y \label{4th-general-0} \\
& =     \int x_0^2 \frac{f_4^2}{f} + (x_1^2+2x_0x_1-2x_0x_3)\frac{f_1^2f_3^2}{f^3} \notag\\
&\indent + (x_2^2-\frac{4}{3}x_0x_2+\frac{4}{3}x_0x_3-\frac{2}{3}x_1x_2) \frac{f_2^4}{f^3}\notag\\
&\indent + (x_3^2+12x_0x_3 - 56x_0x_4 + 4x_1x_3 \notag\\
&\indent \indent -10x_1x_4 +2x_2x_4)\frac{f_1^4f_2^2}{f^5}\notag \\
&\indent + (x_4^2+\frac{240x_0x_4}{7}+\frac{60x_1x_4}{7}+\frac{12x_3x_4}{7})\frac{f_1^8}{f^7} \notag\\
&\indent + (-x_0x_1-4x_0x_2)\frac{f_2f_3^2}{f^2} \notag\\
&\indent + (4x_0x_2-13x_0x_3+12x_0x_4+2x_1x_2 \notag\\
&\indent \indent -3x_1x_3+2x_2x_3)\frac{f_1^2f_2^3}{f^4}\d y, \label{4th-general}
\end{align}
where the terms in \eqref{4th-general-0} are from \eqref{f1f3f4}, \eqref{f22f4}, \eqref{f12f2f4}, \eqref{f14f4}, \eqref{f1f22f3}, \eqref{f13f2f3}, \eqref{f15f3}, and \eqref{f16f2}, respectively.

Therefore,
\begin{align}
  & \int f \left( \frac{f_4}{f} -\frac{6}{5} \frac{f_1f_3}{f^2} -\frac{7}{10} \frac{f_2^2}{f^2} +\frac{8}{5} \frac{f_1^2f_2}{f^3} -\frac{1}{2} \frac{f_1^4}{f^4} \right)^2 \d y \notag\\
& =     \int \frac{f_4^2}{f} + \left((-\frac{6}{5})^2+2(-\frac{6}{5})-2(\frac{8}{5})\right)\frac{f_1^2f_3^2}{f^3}\notag \\
& \indent + \bigg((-\frac{7}{10})^2-\frac{4}{3}(-\frac{7}{10})+\frac{4}{3}(\frac{8}{5}) \notag\\
& \indent  -\frac{2}{3}(-\frac{6}{5})(-\frac{7}{10})\bigg) \frac{f_2^4}{f^3} \notag\\
& \indent + \bigg ((\frac{8}{5})^2+12(\frac{8}{5}) - 56(-\frac{1}{2}) + 4(-\frac{6}{5})(\frac{8}{5}) \notag\\
& \indent  -10(-\frac{6}{5})(-\frac{1}{2}) +2(-\frac{7}{10})(-\frac{1}{2})\bigg)\frac{f_1^4f_2^2}{f^5}  \notag\\
& \indent + \bigg((-\frac{1}{2})^2+\frac{240(-\frac{1}{2})}{7}+\frac{60(-\frac{6}{5})(-\frac{1}{2})}{7} \notag\\
& \indent  +\frac{12(\frac{8}{5})(-\frac{1}{2})}{7}\bigg)\frac{f_1^8}{f^7} \notag\\
& \indent + \left(-(-\frac{6}{5})-4(-\frac{7}{10})\right)\frac{f_2f_3^2}{f^2} \notag \\
& \indent + \bigg(4(-\frac{7}{10})-13(\frac{8}{5})+12(-\frac{1}{2})+2(-\frac{6}{5})(-\frac{7}{10}) \notag\\
&\indent  -3(-\frac{6}{5})(\frac{8}{5})+2(-\frac{7}{10})(\frac{8}{5})\bigg)\frac{f_1^2f_2^3}{f^4}\d y \notag\\
& =\int \frac{f_4^2}{f} - \frac{104f_1^2f_3^2}{25f^3} + \frac{899f_2^4}{300f^3} + \frac{1839f_1^4f_2^2}{50f^5} \notag\\
&\indent -  \frac{1837f_1^8}{140f^7}+ \frac{4f_2f_3^2}{f^2} - \frac{122f_1^2f_2^3}{5f^4} \d y \label{4th-7}
\end{align}

\begin{align}
  & \int f \left( \frac{2}{5} \frac{f_1f_3}{f^2} -\frac{1}{3} \frac{f_1^2f_2}{f^3}  + \frac{9}{100} \frac{f_1^4}{f^4}\right)^2\d y \notag \\
& =  \int \left((\frac{2}{5})^2\right)\frac{f_1^2f_3^2}{f^3} \notag \\
& \indent + \left((-\frac{1}{3})^2 + 4(\frac{2}{5})(-\frac{1}{3}) -10(\frac{2}{5})(\frac{9}{100}) \right)\frac{f_1^4f_2^2}{f^5} \notag \\
& \indent  + \left((\frac{9}{100})^2+\frac{60(\frac{2}{5})(\frac{9}{100})}{7}+\frac{12(-\frac{1}{3})(\frac{9}{100})}{7}\right)\frac{f_1^8}{f^7}  \notag\\
& \indent + \left(-3(\frac{2}{5})(-\frac{1}{3})\right)\frac{f_1^2f_2^3}{f^4}\d y \notag \\
& =  \int \frac{4f_1^2f_3^2}{25f^3} - \frac{704f_1^4f_2^2}{900f^5} + \frac{18567f_1^8}{70000f^7}  + \frac{2f_1^2f_2^3}{5f^4}\d y\label{4th-8}
\end{align}

\begin{align}
& \int f \left( -\frac{4}{100} \frac{f_1^2f_2}{f^3}  + \frac{4}{100} \frac{f_1^4}{f^4}\right)^2\d y \notag\\
&  =     \int  \left((-\frac{4}{100})^2\right)\frac{f_1^4f_2^2}{f^5}\notag \\
& \indent + \left((\frac{4}{100})^2+\frac{12(-\frac{4}{100})(\frac{4}{100})}{7}\right)\frac{f_1^8}{f^7}\d y \notag\\
& =  \int \frac{16f_1^4f_2^2}{10000f^5} - \frac{80f_1^8}{70000f^7}\d y \label{4th-9}
\end{align}

By \eqref{4th-7}, \eqref{4th-8} and \eqref{4th-9},
\begin{align}
& -\frac{1}{2}\int f \left( \frac{f_4}{f} -\frac{6}{5} \frac{f_1f_3}{f^2} -\frac{7}{10} \frac{f_2^2}{f^2} +\frac{8}{5} \frac{f_1^2f_2}{f^3} -\frac{1}{2} \frac{f_1^4}{f^4} \right)^2  \notag\\
& \indent  + f \left(\frac{2}{5} \frac{f_1f_3}{f^2}  -\frac{1}{3} \frac{f_1^2f_2}{f^3} + \frac{9}{100} \frac{f_1^4}{f^4}\right)^2 \notag\\
& \indent  + f \left( -\frac{4}{100} \frac{f_1^2f_2}{f^3}  + \frac{4}{100} \frac{f_1^4}{f^4}\right)^2 \notag\\
& \indent  + \frac{1}{300}   \frac{f_2^4}{f^3} +\frac{56}{90000}  \frac{f_1^4f_2^2}{f^5} + \frac{13}{70000}  \frac{f_1^8}{f^7} \d y \notag\\
& =-\frac{1}{2}\int \bigg(\frac{f_4^2}{f} - \frac{104f_1^2f_3^2}{25f^3} + \frac{899f_2^4}{300f^3} + \frac{1839f_1^4f_2^2}{50f^5}\notag \\
&\indent  -  \frac{1837f_1^8}{140f^7}+ \frac{4f_2f_3^2}{f^2} - \frac{122f_1^2f_2^3}{5f^4}\bigg) \notag \\
&\indent  +\left(\frac{4f_1^2f_3^2}{25f^3} - \frac{704f_1^4f_2^2}{900f^5} + \frac{18567f_1^8}{70000f^7}  + \frac{2f_1^2f_2^3}{5f^4}\right)\notag \\
&\indent  + \left(\frac{16f_1^4f_2^2}{10000f^5} - \frac{80f_1^8}{70000f^7}\right) \notag \\
&\indent  + \frac{1}{300}   \frac{f_2^4}{f^3} +\frac{56}{90000}  \frac{f_1^4f_2^2}{f^5} + \frac{13}{70000}  \frac{f_1^8}{f^7}\d y \notag\\
&= -\frac{1}{2}\int \frac{f_4^2}{f}+ (-\frac{104}{25}+\frac{4}{25} )\frac{f_1^2f_3^2}{f^3} + (\frac{899}{300}+\frac{1}{300}) \frac{f_2^4}{f^3} \notag\\
&\indent  +(\frac{1839}{50}-\frac{704}{900}+\frac{16}{10000} +\frac{56}{90000} ) \frac{f_1^4f_2^2}{f^5} \notag\\
&\indent + (-\frac{1837}{140}+\frac{18567}{70000}-\frac{80}{70000} +  \frac{13}{70000} )\frac{f_1^8}{f^7} \notag \\
&\indent + \frac{4f_2f_3^2}{f^2} +(-\frac{122}{5}+\frac{2}{5}) \frac{f_1^2f_2^3}{f^4} \d y \notag \\
&= -\frac{1}{2}\int \frac{f_4^2}{f}-4\frac{f_1^2f_3^2}{f^3} + 3 \frac{f_2^4}{f^3} +36 \frac{f_1^4f_2^2}{f^5} \notag\\
&\indent   -\frac{90}{7}\frac{f_1^8}{f^7}+ \frac{4f_2f_3^2}{f^2} -24 \frac{f_1^2f_2^3}{f^4} \d y \notag \\
&=\eqref{4thstop-2}, \notag
\end{align}
which completes the proof to Theorem~\ref{mainTheorem-h4}.

\fi

%%%%%%%%%%%%%%%%%%%%%%%%%%%%

\section{Alternative signed representations}\label{sec--13}
In this section, we discuss alternative signed representations of $\frac{\p^n }{\p t^n} h(Y_t)$ in Lemma~\ref{lemma:1}, Theorem~\ref{mainTheorem}, and Theorem~\ref{mainTheorem-h4}.
For the first order derivative, the representation is unique due to its simplicity. For the second and third order derivatives, we have the following alternative representations stated in Corollary~\ref{col:deri-2} and \ref{col:deri-3}.
The proof of Corollary~\ref{col:deri-2}, though simple, contains the idea of how we obtain the formulae in Theorem~\ref{mainTheorem} and \ref{mainTheorem-h4}.
\begin{Corollary}\label{col:deri-2}
\begin{align*}
&\frac{\p^2 }{\p t^2} h(Y_t) \\
&= - \frac12\int f\left(\alpha \frac{f_2}{f} +\beta \frac{f_1^2}{f^2} \right)^2 + f\left(\gamma \frac{f_1^2}{f^2} \right)^2 \\
&\qquad + (1-\alpha^2) \frac{f_2^2}{f} + (-\beta^2 -\gamma^2 -\frac43 \alpha\beta -\frac13) \frac{f_1^4}{f^3}\d y
\end{align*}
where
\begin{align}\label{eqn:coef-2}
\begin{aligned}
1 -\alpha^2 &\geq 0\\
-\beta^2 -\gamma^2 -\frac43 \alpha\beta -\frac13 &\geq 0
\end{aligned}
\end{align}
One set of solution is
\begin{align*}
\alpha = 1,  \quad \gamma = 0, \quad -1\leq \beta \leq -\frac13,
\end{align*}
where the case $\beta = -1$ corresponds to the result in Lemma~\ref{lemma:1}.
\end{Corollary}

\begin{IEEEproof}
After applying the heat equation, the orders of derivatives in each term of $\frac{\p^2 }{\p t^2} h(Y_t)$ have sum equals four. Thus we consider expressing the second derivative as
\begin{align*}
2\frac{\p^2 }{\p t^2} h(Y_t) = - \sum_i \int f\left(\alpha_i \frac{f_2}{f} +\beta_i \frac{f_1^2}{f^2} \right)^2\d y,
\end{align*}
where $\alpha_i$ and $\beta_i$ are coefficients. Since for the reals $A$, $B$, $C$, $a$, $b$, the following equality holds
\begin{align*}
&(aA+B)^2 + (bA+C)^2 \\
&= \left( \sqrt{a^2+b^2} A + \frac{a}{\sqrt{a^2+b^2}} B + \frac{b}{\sqrt{a^2+b^2}} C \right)^2 \\
&\quad + \left( \frac{b}{\sqrt{a^2+b^2}} B - \frac{a}{\sqrt{a^2+b^2}} C \right)^2,
\end{align*}
it suffices to consider the following expression
\begin{align*}
2\frac{\p^2 }{\p t^2} h(Y_t) = - \int f\left(\alpha \frac{f_2}{f} +\beta \frac{f_1^2}{f^2} \right)^2 + f\left(\gamma \frac{f_1^2}{f^2} \right)^2 \d y.
\end{align*}
Now similar to the proof to Theorem~\ref{mainTheorem},
\begin{align*}
&- \int f\left(\alpha \frac{f_2}{f} +\beta \frac{f_1^2}{f^2} \right)^2 + f\left(\gamma \frac{f_1^2}{f^2} \right)^2 \d y \\
& = - \int \alpha^2 \frac{f_2^2}{f} +2\alpha\beta \frac{f_1^2f_2}{f^2} +(\beta^2+\gamma^2) \frac{f_1^4}{f^3} \d y \\
&\eqt{\eqref{eqn:1221}} - \int \alpha^2 \frac{f_2^2}{f} +(\beta^2+\gamma^2+\frac43 \alpha\beta) \frac{f_1^4}{f^3} \d y.
\end{align*}
Comparing with \eqref{eqn:deri-2-final}, one obtains
\begin{align*}
&2\frac{\p^2 }{\p t^2} h(Y_t) \\
&= - \int f\left(\alpha \frac{f_2}{f} +\beta \frac{f_1^2}{f^2} \right)^2 + f\left(\gamma \frac{f_1^2}{f^2} \right)^2 \\
&\qquad + (1-\alpha^2) \frac{f_2^2}{f} + (-\beta^2 -\gamma^2 -\frac43 \alpha\beta -\frac13) \frac{f_1^4}{f^3}\d y.
\end{align*}
To show that the second derivative is negative, one requires
\begin{align*}
1 -\alpha^2 &\geq 0 \\
-\beta^2 -\gamma^2 -\frac43 \alpha\beta -\frac13 &\geq 0.
\end{align*}
And it is easy to verify the set of solution
\begin{align*}
\alpha = 1,  \quad \gamma = 0, \quad -1\leq \beta \leq -\frac13.
\end{align*}

\ifexpand
We have
\begin{align*}
& - \int f \left(\frac{f_2}{f} +\alpha\frac{f_1^2}{f^2}\right)^2 - (\alpha+1) (\alpha+\frac13) \frac{f_1^4}{f^3} \d y \\
&= \int -\frac{f_2^2}{f} -2\alpha \frac{f_1^2f_2}{f^2} +(\frac43 \alpha +\frac13) \frac{f_1^4}{f^3} \d y \\
&\eqt{\eqref{eqn:1221}} \int -\frac{f_2^2}{f} -\frac43 \alpha \frac{f_1^4}{f^3} +(\frac43 \alpha +\frac13) \frac{f_1^4}{f^3} \d y \\
&= \eqref{eqn:deri-2-final}.
\end{align*}
\fi

\end{IEEEproof}

For the third derivative, similar to Corollary~\ref{col:deri-2}, one could determine the coefficients $c_i$ in the following
\begin{align*}
2\frac{\p^3 }{\p t^3} h(Y_t)
&= \int f\left( c_0 \frac{f_3}{f} +c_1 \frac{f_1f_2}{f^2} + c_2 \frac{f_1^3}{f^3}\right)^2 \\
&\quad +f\left( c_3 \frac{f_1f_2}{f^2} + c_4 \frac{f_1^3}{f^3}\right)^2 +f\left(c_5 \frac{f_1^3}{f^3}\right)^2 \d y.
\end{align*}
Since there is no essential difference, we would not present the general expression for the third derivative, but just prove the following corollary.

\begin{Corollary}\label{col:deri-3}
\begin{align*}
\frac{\p^3 }{\p t^3} h(Y_t)
&= \frac12\int f\left( \frac{f_3}{f} - \frac{f_1f_2}{f^2} + \beta \frac{f_1^3}{f^3}\right)^2 \\
&\quad + \left(6\beta - 2\right) \frac{f_1^2f_2^2}{f^3} + (\frac65-\frac{16}{5}\beta -\beta^2) \frac{f_1^6}{f^5}\d y,
\end{align*}
where $\frac{1}{3} \leq \beta \leq  \frac{-8+\sqrt{94}}{5}$.
\end{Corollary}

\begin{IEEEproof}
We have
\begin{align*}
 & \int f\left( \frac{f_3}{f}-\frac{f_1f_2}{f^2}+\beta\frac{f_1^3}{f^3}\right)^2   + (6\beta - 2) \frac{f_1^2f_2^2}{f^3} \\
&\quad  +(\frac65-\frac{16}{5}\beta -\beta^2) \frac{f_1^6}{f^5}  \d y  \\
&= \int \frac{f_3^2}{f} + (6\beta -1) \frac{f_1^2f_2^2}{f^3} +( \frac65-\frac{16}{5}\beta) \frac{f_{1}^6}{f^5} \\
&\quad  - 2\frac{f_1f_2f_3}{f^2}  + 2\beta\frac{f_1^3f_3}{f^3} - 2\beta\frac{f_1^4f_2}{f^4}  \d y  \\
&\eqt{\text{Lemma }\ref{lem:deri-3}}
\int \frac{f_3^2}{f} + (6\beta -1) \frac{f_1^2f_2^2}{f^3} +( \frac65-\frac{16}{5}\beta) \frac{f_1^6}{f^5} \\
&\quad  - 2\left( -\frac{f_2^3}{2f^2} + \frac{f_1^2f_2^2}{f^3} \right)  + 2\beta \left( - \frac{3f_1^2f_2^2}{f^3} +\frac{12f_1^6}{5f^5} \right) \\
&\quad - 2\beta \left( \frac{4f_1^6}{5f^5} \right) \d y \\
&= \int \frac{f_3^2}{f} -3 \frac{f_1^2f_2^2}{f^3} + \frac65 \frac{f_1^6}{f^5} +\frac{f_2^3}{f^2}   \d y \\
& =  \eqref{eqn:deri-3-final}.
\end{align*}
The interval of $\beta$ ensures that the coefficients are positive.
\end{IEEEproof}

For the second and third order derivatives of $h(Y_t)$, the representations can be obtained by hand. For the fourth order derivative, we consider the following representation
\begin{align*}
&2\frac{\p^4 }{\p t^4} h(Y_t) \\
&= -\int f\left( c_0 \frac{f_4}{f} +c_1 \frac{f_1f_3}{f^2} + c_2 \frac{f_2^2}{f^2} + c_3\frac{f_1^2f_2}{f^3} + c_4\frac{f_1^4}{f^4} \right)^2 \\
&\quad +f\left( c_5 \frac{f_1f_3}{f^2} + c_6 \frac{f_2^2}{f^2} + c_7\frac{f_1^2f_2}{f^3} + c_8\frac{f_1^4}{f^4} \right)^2 \\
&\quad +f\left( c_9 \frac{f_2^2}{f^2} + c_{10} \frac{f_1^2f_2}{f^3} + c_{11} \frac{f_1^4}{f^4} \right)^2 \\
&\quad +f\left( c_{12} \frac{f_1^2f_2}{f^3} + c_{13} \frac{f_1^4}{f^4} \right)^2 +f\left(c_{14} \frac{f_1^4}{f^4} \right)^2 \d y.
\end{align*}
By \eqref{eqn:deri-4-general}, we can obtain some constraints similar to \eqref{eqn:coef-2}, and finally find the feasible set of coefficients in Theorem~\ref{mainTheorem-h4} by numerical methods. The process is much more complicated, and we would not present it here.

\section{Conjectures}\label{sec--12}
%\red{This section has been updated!}

Motivated by Theorem \ref{mainTheorem} -- \ref{Costaepi}, we would like to introduce the following conjectures.
\begin{Conjecture}\label{conj1}
The $n$-th order derivative of $h(Y_t)$ satisfies
\begin{itemize}
\item[1.] $\frac{\partial^{n}}{\partial t^n}h(Y_t)\leq 0$ when $n$ is even;
\item[2.] $\frac{\partial^{n}}{\partial t^n}h(Y_t)\geq 0$ when $n$ is odd;
\end{itemize}
i.e., $\frac{\partial^{n}}{\partial t^n}h(Y_t)$ is either convex or concave in $t$ for a  fixed $n$.
\end{Conjecture}
It is easy to see that when $X$ is Gaussian, the above conjectures hold. Conjecture~1 speculates that for a fixed $n$, the convexity or concavity of $\frac{\partial^{n}}{\partial t^n}h(Y_t)$ remains as if $X$ is Gaussian. Conjecture~1 has been verified for $n\leq 2$ in the literature
(Lemma~\ref{lemma:1}), and for $n=3,4$ by Theorem~\ref{mainTheorem} and~\ref{mainTheorem-h4}. 
\begin{Remark} The general pattern for the signed form of the $n$-th order derivative is that, first we need to find all the partitions of $n$, and then each partition is an item in the squares. But the exact coefficients are hard to obtain. One can apply the same technique to deal with the fifth derivative, or even higher. However, the manipulation by hand is huge and hence it is  prohibitive in computational cost, unless one can find some patterns for the coefficients in the signed representations. Some softwares like Mathematica may be useful to verify the higher order derivatives based on the simple rules observed from the fourth derivative, but  we still need a mathematical proof. 
\end{Remark}
The second conjecture is on the log-convexity of Fisher information. From the grand picture of differential entropy and Fisher information, nearly every result on different entropy has a counterpart in Fisher information, e.g., Shannon EPI and FII, the concavity of $h(Y_t)$ and the convexity of $J(Y_t)$ as well as de Bruijn' identity. Corresponding to Costa's EPI, there may be a strengthened convexity of $J(Y_t)$.
\begin{Conjecture}[log-convex]
\label{conj2}
$\log J(Y_t)$ is convex in $t$.
\end{Conjecture}
When $X$ is standard white Gaussian, $J(Y_t)=\frac{1}{t+1}$. We may speculate $\frac{1}{J(Y_t)}$ or $\log J(Y_t)$ is convex in $t$. Simulations show that $\frac{1}{J(Y_t)}$ is neither convex nor concave.
%\blue{Here is an example of $\frac{1}{J(Y_t)}$. Let $X~=~0.5N(0, 0.1)~+~0.5 N(10,0.1)$.  $\frac{1}{J(Y_t)}$ and $\frac{\p^2}{\p t^2}\frac{1}{J(Y_t)}$ are illustrated in Fig. \ref{testJ}.}
Fig.~\ref{testJ} illustrates an example of $\frac{1}{J(Y_t)}$, where $X$ is mixed Gaussian with p.d.f. $g(x)= 0.5f_G(0, 0.1) + 0.5 f_G(10,0.1)$ and $f_G(\mu,\sigma^2)$ is the p.d.f. of Gaussian $\mathcal N(\mu,\sigma^2)$. Limited simulations show that $\log J(Y_t)$ is convex.

\begin{Remark}
After finishing this paper, we realized that Conjecture~\ref{conj1} implies Conjecture~\ref{conj2}. See Section~\ref{sec:FD} for the details.
\end{Remark}

\ifexpand

%\onecolumn
\begin{figure}
  \centering
  \includegraphics[height=300pt,width=400pt]{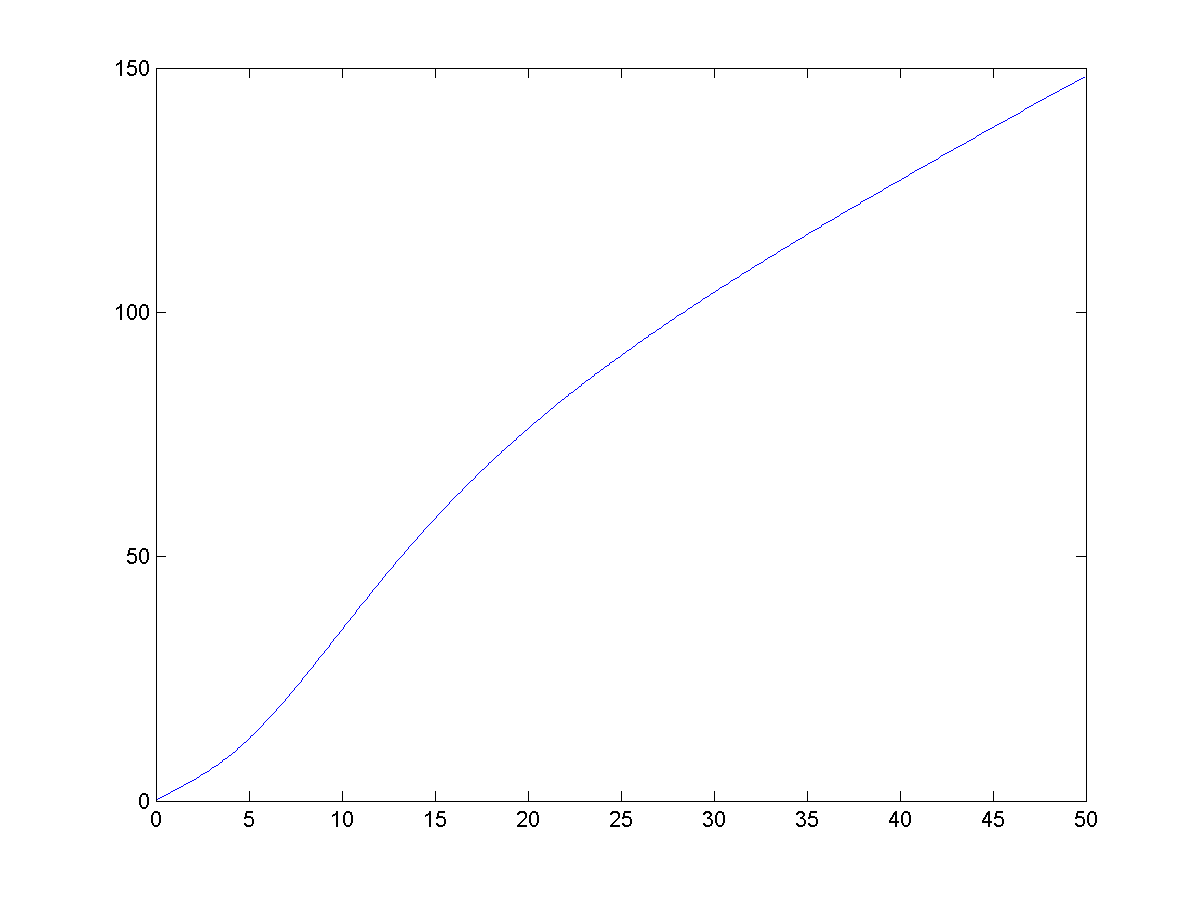}\\
   \caption{$\frac{1}{J(Y_t)}$}\label{testJ}
\end{figure}
\begin{figure}
  \centering
  %\includegraphics[height=300pt,width=400pt]{J-Yt.pdf}\\
   \includegraphics[height=300pt,width=400pt]{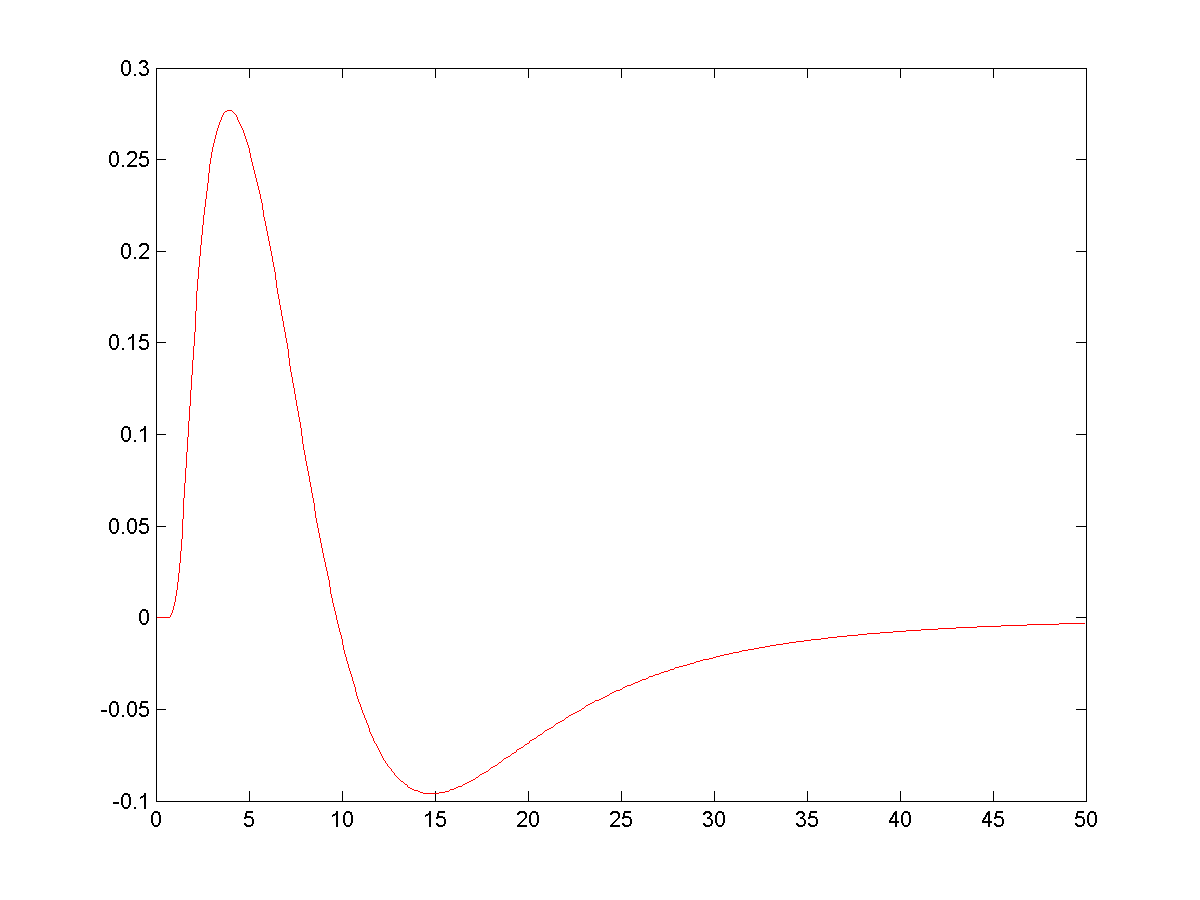}\\
   \caption{$\frac{\p^2}{\p t^2}\frac{1}{J(Y_t)}$}\label{testJ-1}
\end{figure}
%\twocolumn

\fi

\begin{figure}
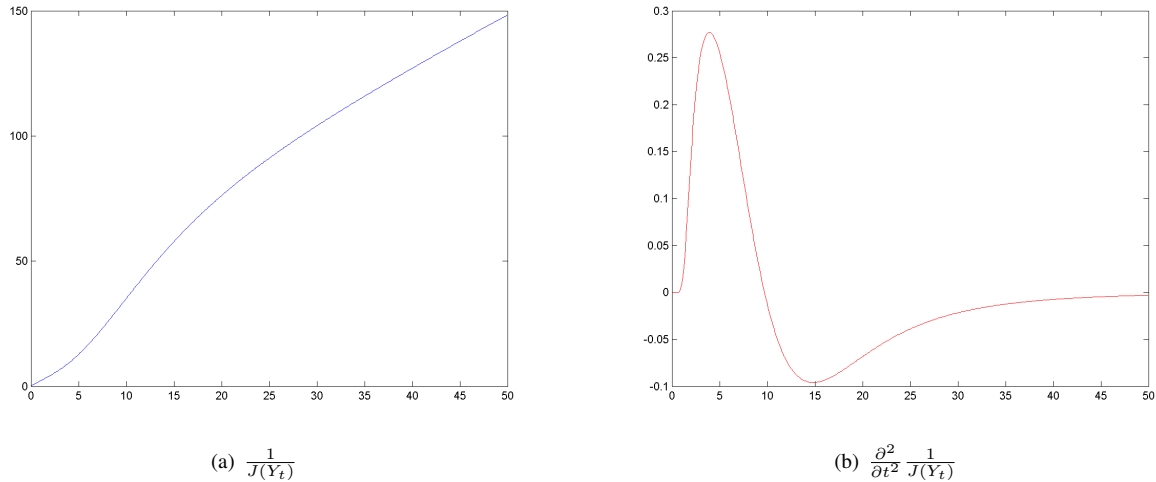
\centering
\subfigure[ $\frac{1}{J(Y_t)}$ ]{
  \includegraphics[width=.45\textwidth]{png-J-Yt.png}
  %\caption{$\frac{1}{J(Y_t)}$}
  \label{Jinv}
}
\subfigure[ $\frac{\p^2}{\p t^2}\frac{1}{J(Y_t)}$ ]{
  \includegraphics[width=.45\textwidth]{png-J-Yt-d2.png}
  %\caption{$\frac{\p^2}{\p t^2}\frac{1}{J(Y_t)}$}
  \label{Jinv-d2}
}
\caption{$\frac{1}{J(Y_t)}$ is neither concave  nor convex.}
\label{testJ}
\end{figure}

% The second conjecture is a strengthened convexity of $J(Y_t)$, which can be seen   Since $h(Y_t)$ is exponentially concave and there are many similar correspondences between Fisher information and differential entropy, e.g., FII and EPI, it is reasonable to speculate that there is a counterpart of Costa's EPI for Fisher information; i.e., $\log J(Y_t)$ is convex. Also, when $X$ is Gaussian, we have $J(X+\sqrt{t}Z) = \frac{1}{t+a}$. In principle, if certain stronger convexity exists for $J(Y_t)$, then $\log J(Y_t)$ is a potential candidate. (Note that $1/J(Y_t)$ is not convex.)

\section{Concavity of $h(\sqrt{t}X + \sqrt{1-t}Z)$} \label{sec--14}
%\blue{
%Likewise, we define the $n$-th order information as
%\begin{equation}
%I_n(X) := \frac{\partial^{n}}{\partial t^n}h(Y_t)\bigg |_{t=0}.
%\end{equation}
%Then $I_0(X) = h(X)$ and $I_1(X) = \frac12 J(X)$.
%Let
%\begin{equation}
%W_t := \sqrt{t}X + \sqrt{1-t}Z.
%\end{equation}
%In this section, we study the convexity or concavity of $I_n(W_t)$.
%}

%\red{ Suggestion: using $t$ for $W_t$ introduces misunderstanding compared with $Y_t$, since they have different intervals. May use $\lambda$ or $s$ instead.}

For $0<t<1$,  let
\begin{equation}
W_t := \sqrt{t}X + \sqrt{1-t}Z,
\end{equation}
where $Z\sim\mathcal N(0,1)$ is independent of $X$. In this section, we study the  concavity and convexity  of $h(W_t)$ and $J(W_t)$, respectively.

Lieb showed that Shannon EPI (\ref{ShannonEPI}) is equivalent to
\begin{equation}\label{Lieb}
h(\sqrt{\lambda} X_1+ \sqrt{1-\lambda} X_2)\geq \lambda h(X_1) +  (1-\lambda) h(X_2)
\end{equation}
for any $0 \leq \lambda \leq 1$. Here we use $X_1$ and $X_2$ in lieu of $X$ and $Y$ as the independent random variables.

In the literature, $(X_1, X_2)\to \sqrt{\lambda} X_1+ \sqrt{1-\lambda} X_2$ is referred to as the covariance-preserving transformation, which can be found in many generalizations of Shannon EPI (\cite{Rioul11}). The original proof of Lieb is a little tricky. Next, we give a geometrical interpretation of this transformation which can help us to have a better appreciation on    $\sqrt{\lambda} X_1+ \sqrt{1-\lambda} X_2$.
\subsection{Covariance-preserving Transformation}
%\subsection{Lieb's EPI}
%First, we state some results on convex optimization for further study.
Recall that a convex function has the following three equivalent statements. %\red{these are not definitions, but equivalent statements}

Let $f(x)$ be a function which is twice differentiable, where $x\in \mathbb{R}^n$. Then the following are equivalent:
\begin{itemize}
\item [1.] $f(x)$ is convex in $x$.

\item [2.] The Hessian matrix of  $f(x)$ is positive {semi-definite}; i.e.,
\begin{equation}
    \nabla^2 f\succeq 0.
\end{equation}

\item [3.] For any fixed point $x_0$,
\begin{equation}\label{eqn:convex1}
    f(x)\geq f(x_0) +(x-x_0)^T\nabla f(x_0).
\end{equation}
\end{itemize}

Furthermore, $y=f(x_0) +(x-x_0)^T\nabla f(x_0)$ can be viewed as the tangent plane at point $(x_0,f(x_0))$  {for function $y=f(x)$}.
In the following, we shall apply the above argument on convex functions to study the so-called covariance-preserving transformation.

Shannon EPI (\ref{ShannonEPI}) can be equivalently transformed to
\begin{equation}
h(X_1+X_2) \geq \frac12\log \left( e^{2h(X_1)}+e^{2h(X_2)} \right).
\end{equation}
Let's study function $f(x_1, x_2) =\frac{1}{2}\log\left( e^{2x_1}+e^{2x_2}\right)$.
%\blue{
%It is easy to see that $f(x_1, x_2)$ is convex in $(x_1,x_2)$ by $\nabla^2~f(x_1,x_2)~\succeq~0$. By some algebra,
%\begin{align}
%\nabla f = (\frac{\partial f} {\partial x_1},\frac{\partial f} {\partial x_2}) = (\frac{e^{2x_1}}{e^{2x_1}+e^{2x_2}},   \frac{e^{2x_2}}{e^{2x_1}+e^{2x_2}}).
%\end{align}
%}
{
By some manipulations,
\begin{align*}
\nabla f &= \left(\frac{\partial f} {\partial x_1},\frac{\partial f} {\partial x_2}\right) = \left(\frac{e^{2x_1}}{e^{2x_1}+e^{2x_2}}, \frac{e^{2x_2}}{e^{2x_1}+e^{2x_2}}\right), \\
\nabla^2 f &=
\left[ \frac{\p^2 f}{\p x_ix_j} \right]_{ij}
 = \frac{2e^{2x_1}e^{2x_2}}{(e^{2x_1}+e^{2x_2})^2} \begin{bmatrix}
 1 & -1 \\
 -1 & 1
 \end{bmatrix}.
\end{align*}
It is easy to see that $f(x_1, x_2)$ is convex since $\nabla^2 f \succeq 0$.
}
By \eqref{eqn:convex1}, the tangent plane of  $f(x_1,x_2)$ at point $(x_1,x_2)=(\frac{1}{2}\log(\sigma_1^2),\frac{1}{2}\log(\sigma_2^2))$ is
\begin{align}
y & = \frac{1}{2}\log\left(\sigma_1^2+\sigma_2^2\right) + \left(x_1 -\frac{1}{2}\log(\sigma_1^2) \right)\frac{\sigma_1^2}{\sigma_1^2  + \sigma_2^2} \notag\\
  &  + \left(x_2 -\frac{1}{2}\log(\sigma_2^2) \right)\frac{\sigma_2^2}{\sigma_1^2  + \sigma_2^2}.
\end{align}
Hence, (\ref{ShannonEPI}) is equivalent to
\begin{align}
h(X_1+X_2)& \geq  \frac12 \log\left(\sigma_1^2+\sigma_2^2\right) + \left(h(X_1) -\frac{1}{2}\log(\sigma_1^2) \right) \frac{\sigma_1^2}{\sigma_1^2  + \sigma_2^2} \notag\\
& \indent + \left(h(X_2) -\frac{1}{2}\log(\sigma_2^2) \right) \frac{\sigma_2^2}{\sigma_1^2  + \sigma_2^2}.
\end{align}
Let
$$\lambda = \frac{\sigma_1^2}{\sigma_1^2  + \sigma_2^2}.$$
%\red{$\lambda = \sigma_1^2/ (\sigma_1^2  + \sigma_2^2)$}.
%\blue{By some algebra,}
{Notice that $h(aX) = h(X) +\log|a|$,}
we have
\begin{equation}\label{eqn:tangent}
h(X_1+X_2) \geq \lambda h(X_1/\sqrt{\lambda}) + (1- \lambda) h(X_2/\sqrt{1-\lambda}).
\end{equation}
Substitute $(X_1, X_2)$ with $(\sqrt{\lambda}X_1, \sqrt{1-\lambda}X_2)$,
\begin{equation}
h(\sqrt{\lambda} X_1+\sqrt{1- \lambda}X_2) \geq \lambda h(X_1) + (1- \lambda) h(X_2),
\end{equation}
which is exactly the inequality (\ref{Lieb}).

In the above proof, the points   share the same tangent plane \eqref{eqn:tangent} as long as they admit the same $\lambda$. In fact, all the results (see \cite{Rioul11}) that applied  covariance-preserving transformation can be proved in this manner.

\subsection{The concavity of $h(W_t)$}
\begin{Theorem}\label{them-hwt}
$h(W_t)$ is concave in $t$, $0 <  t < 1$.
\end{Theorem}
\begin{IEEEproof}
Since
$$h(W_t)= h(X + \sqrt{1/t-1}Z)+\frac{1}{2}\log t,$$
by some algebra, we obtain
\begin{align}
\frac{\p}{\p t} h(W_t) = \frac{1}{2}J(X + \sqrt{1/t-1}Z) \left(-\frac{1}{t^2}\right) + \frac{1}{2t}
\end{align}
and
\begin{align}
\frac{\p^2}{\p t^2} h(W_t) &= \frac{1}{2}J'(X + \sqrt{1/t-1}Z) \left(-\frac{1}{t^2}\right)^2 \notag \\
&\quad + \frac{1}{2}J(X + \sqrt{1/t-1}Z) \left(\frac{2}{t^3}\right) - \frac{1}{2t^2}.
\end{align}
To show
$$\frac{\p^2}{\p t^2} h(W_t)\leq 0,$$
we need to prove
\begin{align*}
&-\frac{1}{2}J'(X + \sqrt{1/t-1}Z) \left(-\frac{1}{t^2}\right)^2  + \frac{1}{2t^2} \\
&\geq \frac{1}{2}J(X + \sqrt{1/t-1}Z) \left(\frac{2}{t^3}\right).
\end{align*}
That is
\begin{equation}
-J'(X + \sqrt{1/t-1}Z)  + t^2 \geq 2tJ(X + \sqrt{1/t-1}Z). \label{txz-1}
\end{equation}
By \eqref{fisher-costa}, Costa's EPI is equivalent to
\begin{equation}
-J'(X+\sqrt{s}Z)\geq J(X+\sqrt{s}Z)^2, \notag
\end{equation}
for any $s>0$.
Therefore,
%\blue{
%\begin{align*}
%&-J'(X + \sqrt{1/t-1}Z)  + t^2 \\
%& \geq 2\sqrt{-J'(X + \sqrt{1/t-1}Z) \times t^2 }\notag\\
%& \geq 2\sqrt{J(X + \sqrt{1/t-1}Z)^2 \times t^2 }\notag\\
%& = 2tJ(X + \sqrt{1/t-1}Z),
%\end{align*}
%}
{
\begin{align*}
&-J'(X + \sqrt{1/t-1}Z)  + t^2 \\
& \geq J(X + \sqrt{1/t-1}Z)^2  + t^2\\
& \geq 2tJ(X + \sqrt{1/t-1}Z),
\end{align*}
}
which is \eqref{txz-1}.
\end{IEEEproof}
 In all the results above, as $t>0$, $X+\sqrt{t}Z$ can be replaced by $X'+\sqrt{s}Z\big |_{s=0}$, where $X' = X+\sqrt{t}\hat{Z}$ and $\hat{Z}$ is the standard Gaussian and is independent of $X$ and $Z$.
In this manner, we only need to prove that the result holds for any such $X'$ at point $s=0$.   In light of the smoothness introduced by $\sqrt{t}Z$ where $t>0$, without weakening our result, we can just assume that when $t\to 0$, the $n$-th order derivative of  $h(X+\sqrt{t}Z)$ exists in the sequel.

Next, we show that Theorem \ref{them-hwt} can imply Costa's EPI. In the above proof, if $J(X)$ and $J'(X)$ are well defined, then let $t\to 1$ in  \eqref{txz-1},
\begin{align}
-J'(X) +1 \geq 2J(X). \label{g-1}
\end{align}
Let $\hat{X} = X/\sqrt{J(X)}$, then
\begin{equation}
J'(\hat{X}) = \frac{J'(X)}{J(X)^2}, \text{ and  } J(\hat{X}) = 1.
\end{equation}
Substitute $X$ with $\hat{X}$ in \eqref{g-1},
\begin{equation}
-\frac{J'(X)}{J(X)^2} \geq 1,
\end{equation}
which is just Costa's EPI by \eqref{fisher-costa}.

\subsection{The convexity of  $J(W_t)$ }
%\red{It would be better to clearly state the results in the form of, say a Claim.}
In this section, we study  the convexity of $J(W_t)$ via the relations among the convexities of $J(W_t)$, $\frac{1}{J(Y_t)}$, and $\log J(Y_t)$.
\begin{Claim}
$J(W_t)$ is not convex.
\end{Claim}
By some algebra,  $\log J(Y_t)$ is convex in $t$  if and only if
\begin{align} \label{log-fisher}
J''(Y_t)J(Y_t) \geq (J'(Y_t))^2.
\end{align}
$\frac{1}{J(Y_t)}$ is convex in $t$ if and only if
\begin{equation} \label{1-fisher}
J''(Y_t)J(Y_t) \leq 2(J'(Y_t))^2,
\end{equation}
and concave if and only if
\begin{align} \label{eqn:invJ-conc-cond}
J''(Y_t)J(Y_t) \geq 2(J'(Y_t))^2.
\end{align}

The first and second order derivatives of $J(W_t)$ are
\begin{align}
 &   \frac{\p}{\p t} J(W_t) \notag\\
 & = \frac{\p}{\p t} \frac{1}{t}J(X + \sqrt{1/t-1}Z) \notag\\
 & =  -\frac{1}{t^2}J(X + \sqrt{1/t-1}Z) -\frac{1}{t^3}J'(X + \sqrt{1/t-1}Z)
\end{align}
and
\begin{align}
 &   \frac{\p^2}{\p t^2} J(W_t) \notag\\
 & = \frac{2}{t^3}J(X + \sqrt{1/t-1}Z) + \frac{1}{t^4}J'(X + \sqrt{1/t-1}Z) \notag\\
 & \indent + \frac{3}{t^4}J'(X + \sqrt{1/t-1}Z) +  \frac{1}{t^5}J''(X + \sqrt{1/t-1}Z) \notag\\
 & = \frac{2}{t^3}J(X + \sqrt{1/t-1}Z) + \frac{4}{t^4}J'(X + \sqrt{1/t-1}Z) \notag \\
 & \quad +  \frac{1}{t^5}J''(X + \sqrt{1/t-1}Z).\label{fisher-convex}
\end{align}
If we can show that %\red{(here can refer to equation \eqref{eqn:invJ-conc-cond})}
\begin{align}
J''(X+\sqrt{s}Z)J(X+\sqrt{s}Z) \geq 2(J'(X+\sqrt{s}Z))^2 \label{log-fisher-1}
\end{align}
holds for any  $s > 0$,
then  \eqref{log-fisher} holds and
\begin{align}
& \frac{\p^2}{\p t^2} J(W_t) \notag\\
& = \frac{2}{t^3}J(X + \sqrt{1/t-1}Z) + \frac{4}{t^4}J'(X + \sqrt{1/t-1}Z) \notag \\
&\quad +  \frac{1}{t^5}J''(X + \sqrt{1/t-1}Z)\notag\\
& \geq 2\sqrt{\frac{2}{t^3}J\times \frac{1}{t^5}J''} + \frac{4}{t^4}J' \notag\\
& \geq 2\sqrt{\frac{2}{t^8}2(J')^2} + \frac{4}{t^4}J' \notag\\
& \geq 0.
\end{align}

Conversely, if  $\frac{\p^2}{\p t^2} J(W_t) \geq 0$ holds, we can show that \eqref{log-fisher-1} also holds. In \eqref{fisher-convex}, let $t\to 1$, we obtain that
\begin{equation}
2J(X) + 4J'(X) + J''(X)\geq 0.\notag
\end{equation}
Substitute $X$ by $X' = aX$, where $a>0$,
\begin{equation}
2\frac{J(X)}{a^2} + 4\frac{J'(X)}{a^4} + \frac{J''(X)}{a^6}\geq 0. \label{log-fisher-2}
\end{equation}
Note that $J\geq 0$ and $J''\geq 0$. Choose proper $a$ such that
\begin{align}
2\frac{J(X)}{a^2} =  \frac{J''(X)}{a^6}.\notag
\end{align}
Hence
\begin{align}
&2\frac{J(X)}{a^2} =  \frac{J''(X)}{a^6} \notag\\
&= \sqrt{2\frac{J(X)}{a^2} \times  \frac{J''(X)}{a^6}} = \frac{1}{a^4}\sqrt{2J(X)J''(X)}.\notag
\end{align}
Therefore, \eqref{log-fisher-2} becomes
\begin{align}
&2\frac{J(X)}{a^2} + 4\frac{J'(X)}{a^4} + \frac{J''(X)}{a^6} \notag\\
&=  \frac{2}{a^4}\sqrt{2J(X)J''(X)}+ 4\frac{J'(X)}{a^4} \geq 0,\notag
\end{align}
which is
\begin{align}
J(X)J''(X)\geq 2(J'(X))^2.\notag
\end{align}
Hence, $\frac{\p^2}{\p t^2} J(W_t)\geq 0$ if and only if \eqref{log-fisher-1} holds for arbitrary $X$. Because  $\frac{1}{J(Y_t)}$ is neither convex nor concave for arbitrary $X$,
{ which means neither \eqref{1-fisher} nor \eqref{eqn:invJ-conc-cond} holds always, thus $J(W_t)$ is  neither convex nor concave.}

%Because $\log J(Y_t)$ is strictly convex, the corresponding convex condition \eqref{log-fisher} is not as sharp as that \eqref{log-fisher-1} of $J(W_t)$.

\section{Further Discussion}\label{sec:FD}
After the third and fourth derivatives are obtained, we consult  the literature to find more connections and implications. The first finding is that in the literature of mathematical physics, Conjecture 1 was studied in a 1966 paper~\cite[Sec. 12]{McKean1966} by McKean, who studied the signs of the third and fourth derivatives  but failed to prove them. By this means, our results provide an affirmative answer to McKean's problem up to the fourth order. Furthermore,  following the routine rules obtained in our paper, one may try to verify the conjecture up to any finite order. McKean's work has many other conjectures regarding thermodynamics and has remained unknown to information theory community until very recently. For more details on McKean's work, one may refer  to Villani~\cite[pp. 165-166]{VillaniReview}.

Another finding is that  Conjecture 1 can be discussed in the context of completely monotone functions (Widder \cite{WidderLap}).
\begin{Definition}
A function $f(t), t \in (0, \infty)$ is  completely monotone, if for all $n = 0$, $1$, $2$, ..., 
\begin{equation}\notag
(-1)^n \frac{\d^n }{\d t^n} f(t) \geq 0.
\end{equation}
\end{Definition}
Hence, Conjecture 1 can be restated as: 
\begin{Conjecture*}[Completely Monotone Conjecture]
$J(Y_t)$ is completely monotone in $t\in (0, \infty)$.
\end{Conjecture*}
A very interesting result on completely monotone functions is due to Fink~\cite{Fink1982}: If $f(t)$ is completely monotone in $t$, then $f(t)$ is log-convex. By this means, Conjecture 1 can imply Conjecture 2. 

Another result on completely monotone functions is the following theorem (Widder \cite[p. 160]{WidderLap}).
\begin{Theorem}[Bernstein's theorem] 
A necessary and sufficient condition that $f(t)$ should be completely monotone in $[0, \infty)$ is that  
\begin{equation}\notag
f(t) = \int_{0}^{\infty} e^{-tx} \d \alpha(x),
\end{equation}
where $\alpha(x)$ is bounded and non-decreasing and the above integral converges for $0\leq t <\infty$.
\end{Theorem} 
That is, if Conjecture 1 is true,  an equivalent expression for Fisher information will be obtained. Noting that $\alpha (x)$ can be regarded as a measure defined on $[0, \infty)$.

In this paper, to simplify the problem, we  consider only the univariate case of random variables. For the  multivariate case,  the computation will be much more involved. Some sophisticated techniques that have been developed in probability theory may be useful; e.g.,  the $\Gamma_2$ calculus, which can be found in Villani \cite{VillaniEPI} and Bakry \textit{et al}. \cite{BGM2014}. 
 
%%%%%%%%%%%%%%%%%
\section{Conclusion}\label{sec:3}
The Gaussian random variables have many fascinating properties. In this paper, we have obtained the third and the fourth order derivatives of $h(X+\sqrt{t}Z)$. The signed representations have a very interesting form. We wish to show that, though we cannot obtain a closed-form expression on $h(X+\sqrt{t}Z)$ when $X$ is arbitrary, we can still obtain its convexity or concavity for any order derivative. Our progress verifies a small part of the conjectures and has nearly exhausted the power of fundamental calculus. A new approach  may be needed towards solving these conjectures.

In the literature, the approach that employed heat equation and integration by parts is merely one of many different approaches to prove Costa's EPI. For the approaches like data processing argument in \cite{ZamirFII} \cite{Rioul11}, and the advanced tools in \cite{VillaniEPI}, it is unknown whether they can go further than what we have done. However, if these conjectures are correct, a rather fundamental fact about the Gaussian random variable will be revealed in the language of differential entropy.

%%%%%\begin{IEEEbiography}[{\includegraphics[width=1in,height=1.25in,clip,keepaspectratio]{mshell}}]{Michael Shell}
%\begin{IEEEbiographynophoto}{\bf Fan Cheng}(S'12-M'14)
%received the bachelor degree in
%computer science from Shanghai Jiao Tong
%University in 2007, and the PhD degree in
%information engineering from The Chinese
%University of Hong Kong in 2012. As of 2012,
%he has been a postdoctoral fellow in the Institute of
%Network Coding.

\appendix
\subsection{Proof to Proposition~\ref{prop-int-zero}} \label{apdx:prop-int-zero}
The technique used in this proof is essentially the same as that by Costa. One may refer to \cite{CostaEPI} for more details.

\begin{IEEEproof}
One can obtain the formulae for the derivatives as
\begin{align}
f^{(n)}(y,t) = \int g(x)\frac{1}{\sqrt{2\pi t}}e^{-\frac{(y-x)^2}{2t}}H_n(y-x)\d x,\notag
\end{align}
where $H_0 = 1$ and $H_n$ satisfies the recursion formula
\begin{align}
H_n(y-x) = -\frac{y-x}{t} H_{n-1} + \frac{\p}{\p y} H_{n-1}.\notag
\end{align}
In general $H_n$ can be expressed as
\begin{align}
H_n(y-x) = \sum_{j=0}^n\alpha_{n,j}(y-x)^{n-j}\notag
\end{align}
where $\alpha_{n,j}$'s are some constants that also depend on $t$ (and actually these constants are zeroes for odd $j$).

Notice that
\begin{align}
&\frac{f^{(n)}}{f} = \int g(x)\frac{1}{\sqrt{2\pi t}}e^{-\frac{(y-x)^2}{2t}}H_n(y-x) \frac{1}{f(y,t)}\d x \notag\\
&= \e \left[ H_n(Y_t-X)|Y_t =y \right] \notag\\
& = \sum_{j=0}^n \alpha_{n,j} \e \left[ (Y_t-X)^{n-j}|Y_t =y \right].\notag
\end{align}

Let $\alpha_n := \sum_l |\alpha_{n,l}|.$ We prove Proposition~\ref{prop-int-zero} by induction on $r$.
When $r =1$,
\begin{align}
&\int f\left|\frac{f^{(n)}}{f}\right|^k\d y = \e \left[ \left|\frac{f^{(n)}}{f}\right|^{k} \right] \notag\\
&= \e\left[ \left|\sum_{j=0}^n \alpha_{n,j} \e \left[ (Y_t-X)^{n-j}|Y_t =y \right]\right|^{k} \right]\notag\\
&= \alpha_n^{k} \e\left[ \left|\sum_{j=0}^n \frac{\alpha_{n,j}}{\alpha_n}
\e \left[ (Y_t-X)^{n-j}|Y_t =y \right]\right|^{k} \right]  \notag\\ %\label{eqn:coeff-alpha}
&\leq \alpha_n^{k} \sum_{j=0}^n \frac{|\alpha_{n,j}|}{\alpha_n} \e \left[
\left| \e \left[ |Y_t-X|^{n-j}|Y_t =y \right] \right|^{k} \right] \label{eqn:Jensen-comb}\\
&\leq \alpha_n^{k} \sum_{j=0}^n \frac{|\alpha_{n,j}|}{\alpha_n} \e \left[
 \e \left[ |Y_t-X|^{k(n-j)}|Y_t =y \right]  \right] \label{eqn:Jensen-swap}\\
&= \alpha_n^{k} \sum_{j=0}^n \frac{|\alpha_{n,j}|}{\alpha_n}
\e \left[ |Y_t-X|^{k(n-j)}\right]  \notag\\
&= \alpha_n^{k} \sum_{j=0}^n \frac{|\alpha_{n,j}|}{\alpha_n}
\e \left[ \left| \sqrt{t}Z \right|^{k(n-j)}\right] \notag \\
&< +\infty, \notag
\end{align}
where  \eqref{eqn:Jensen-comb} and \eqref{eqn:Jensen-swap} are due to Jensen's inequality.

When $r\geq 2$, by induction,
\begin{align}
& \int f \left|\prod_{i=1}^{r} \frac{[f^{(m_i)}]^{k_i}}{f^{k_i}}\right| \d y = \e\left[ \prod_{i=1}^{r}\left|\frac{f^{(m_i)}}{f} \right|^{k_i} \right] \notag\\
& \leq  \left(\e\left[ \prod_{i=1}^{r-1}\left|\frac{f^{(m_i)}}{f} \right|^{2k_i}\right] \cdot \e\left[ \left|\frac{f^{(m_r)}}{f} \right|^{2k_r}\right]\right)^{\frac12}\label{eqn:inter2} \\
& < +\infty, \notag
\end{align}
where \eqref{eqn:inter2} is by the Cauchy-Schwartz inequality.

The fact that
$f \prod_{i=1}^{r} \frac{[f^{(m_i)}]^{k_i}}{f^{k_i}}$
vanishes as $|y|\to \infty$ can be obtained from the existence of integral.
\end{IEEEproof}

%\subsection{Proof to Lemma~\ref{lemma:heat}} \label{proof-lemma-heat}
%\begin{IEEEproof}
%By some calculus,
%\begin{align*}
% f_t &=  \int g(x)  \frac{1}{\sqrt{2\pi t}} e^{-\frac{(y-x)^2}{2t}}  \left(\frac{(y-x)^2}{2}\frac{1}{t^2}-\frac{1}{2 t}\right) \d x,  \\
% f_y     &=   \int g(x)\frac{1}{\sqrt{2\pi t}}e^{-\frac{(y-x)^2}{2t}}\left(-\frac{1}{t}(y-x)\right) \d x, \\
% f_{yy}  &= \int g(x)\frac{1}{\sqrt{2\pi t}}e^{-\frac{(y-x)^2}{2t}}\left[\left(\frac{1}{t}(y-x)\right)^2-\frac{1}{t}\right] \d x.
%\end{align*}
%By comparing $f_{yy}$ with $f_t$, the lemma can be proved.
%\end{IEEEproof}

\ifexpand

\subsection{Proof to Lemma~\ref{lemma:1}}\label{proof-lemma1}
Here we present the proof of Lemma \ref{lemma:1}, as some intermediate steps are instrumental in the proof of Theorem \ref{mainTheorem}.
\begin{IEEEproof}
For the first order derivative we have
\begin{align}
&\frac{\partial }{\partial t} h(Y_t)  =   \frac{\p}{\p t}\left[-\int f(y,t)\log f(y,t)\d y\right]\notag\\
&=  -\int f\frac{1}{f} f_t \d y -\int f_t \log f \d y\notag\\
&\eqt{\eqref{heatEqn}} -\frac{\p }{\p t}\int f \d y -\int \frac12  f_{yy} \log f \d y \notag\\ % \quad \text{ (by \eqref{heatEqn}) }
&= 0 -\frac12 \int \log f \d f_y\notag\\
&= -\frac12 (\log f) f_{y} \bigg |_{y=-\infty}^{+\infty}  + \frac12 \int f_{y} \d \log f \label{limit:001}\\
&=  0+ \frac{1}{2} \int \frac{f^2_y}{f}\d y \notag\\
&=  \frac{1}{2}J(Y_t).\notag
\end{align}
The limits in  \eqref{limit:001} are zero, because $f_{y}\log f= \frac{f_{y}}{\sqrt{f}} \sqrt{f}\log f$, where  $\frac{f_y^2}{f}\to 0$ from Proposition~\ref{prop-int-zero}, and $x\log x \to 0$, as $x\to 0$.

For the second order derivative, using integration by parts:
\begin{align}
&\frac{\p^2 }{\p t^2}h(Y_t) = \frac{1}{2} \int  \frac{2f_y f_{yt}}{f}  - \frac{f^2_yf_t}{f^2} \d y \notag\\
&= \frac{1}{2} \int \frac{2f_y }{f} \d f_t  - \frac{1}{2}\int\frac{f^2_yf_t}{f^2}\d y \notag\\
&\eqt{\eqref{heatEqn}} \frac{f_tf_y}{f}\bigg |_{y=-\infty}^{+\infty}- \frac{1}{2}\int   \left(\frac{2f_{y}}{f}\right)_yf_t\d y - \frac{1}{2}\int  \frac{f^2_yf_{yy}}{2f^2}\d y \label{limit:002}\\
&= -\frac{1}{2}\int\left(  \frac{f_{yy}^2}{f} - \frac{f_y^2f_{yy}}{2f^2}  \right) \d y. \label{eq31} %+ \frac{f_y^2f_{yy}}{4f^2}
\end{align}
For the last term we have
\begin{align}
&\int \frac{f_y^2f_{yy}}{2f^2} \d y = \int \frac{f_y^2}{2f^2} \d f_y \notag\\
& = \frac{f_y^3}{2f^2}\bigg |_{y=-\infty}^{+\infty}-\int \left(\frac{f_y^2}{2f^2}\right)_y f_y \d y \label{limit:003}\\
&                                   = -\int \frac{f_y^2f_{yy}}{f^2}\d y +\int \frac{f_y^4}{f^3} \d y \notag
\end{align}
Hence,
\begin{align}
\int \frac{3f_y^2 f_{yy}}{2f^2}\d y =\int \frac{f_{y}^{4}}{f^3} \d y. \label{eq32}
\end{align}
The limits in \eqref{limit:002} and \eqref{limit:003} vanish by Proposition \ref{prop-int-zero}.
By \eqref{eq31} and \eqref{eq32},
\begin{align}
&\frac{\p^2 }{\p t^2}h(Y_t) =-\frac{1}{2} \int \frac{f_{yy}^2}{f} - \frac{2f_y^2 f_{yy}}{f^2} + \frac{f_y^4}{f^3} \d y \notag\\
&=  -\frac{1}{2} \int f \left(  \frac{f_{yy}^2}{f^2} - \frac{2f_y^2 f_{yy}}{f^3} + \frac{f_y^4}{f^4}  \right)\d y \notag\\
&=-\frac{1}{2}\int f \left(\frac{f_{yy}}{f}-\frac{f_y^2}{f^2}\right)^2\d y,\notag
\end{align}
and the proof is finished.
\end{IEEEproof}

\fi

\subsection{Proof to Costa's EPI}
\begin{IEEEproof}
Costa's EPI is equivalent to
\begin{equation}
    \frac{\p^2 }{\p t^2}e^{2h(Y_t)} \leq 0.
\end{equation}
By some algebra, one needs to show
\begin{equation}
2 \left(\frac{\p }{\p t} h(Y_t) \right)^2\leq -\frac{\p^2 }{\p t^2} h(Y_t),
\end{equation}
or
\begin{equation}
 J(Y_t)^2\leq -J'(Y_t), \label{fisher-costa}
\end{equation}
i.e.,
\begin{align}
 \int f \left( \frac{f_{yy}}{f} - \frac{f_y^2}{f^2} \right)^2\d y \geq \left( \int \frac{f_y^2}{f}\d y \right)^2,
\end{align}
which can be proved by the inequality of arithmetic and geometric means:
\begin{align}
&\int f \left(\frac{f_{yy}}{f} - \frac{f_y^2}{f^2} \right)^2 \d y  \geq  \left( \int f \left(\frac{f_{yy}}{f} - \frac{f_y^2}{f^2}\right) \d y\right)^2 \notag \\
                                                    & = \left(   \int f_{yy} \d y -  \int \frac{f_y^2}{f} \d y\right)^2\label{fyy0} \\
                                                    & = \left( \int \frac{f_y^2}{f} \d y\right)^2. \notag
\end{align}
In \eqref{fyy0}, $\int f_{yy} \d y = \int 2f_{t} \d y= \frac{\p}{\p t} \int 2f \d y = \frac{\p}{\p t} 2 = 0$.
\end{IEEEproof}

\subsection{Proof to Lemma~\ref{lem:deri-3}} \label{app:lem-deri-3}

\begin{IEEEproof}
We use integration by parts to eliminate the high-order terms:
\begin{align}
\int \frac{f_{1}^4f_{2}}{f^4} \d y
&= \int  \frac{f_{1}^4}{f^4}\d f_1 \notag \\
&= \int  \frac{1}{5f^4} \d f_1^5 \notag \\
&= \frac{f_1^5}{5f^4}\bigg |_{y=-\infty}^{+\infty} -\int \frac{f_1^5}{5} \left(\frac{1}{f^4}\right)_y \d y \notag \\
&\eqt{\eqref{lim:term-zero}} 0 + \int \frac{4f_1^6}{5f^5} \d y . \label{eqn:app-1241}
\end{align}

\begin{align*}
\int \frac{f_1^3f_3}{f^3} \d y
&= \int \frac{f_1^3}{f^3} \d f_2 \\
&= \frac{f_1^3f_2}{f^3} \bigg |_{y=-\infty}^{+\infty} -\int f_2 \left(\frac{f_1^3}{f^3}\right)_y \d y \\
&\eqt{\eqref{lim:term-zero}} 0 -\int f_2 \frac{3f_1^2}{f^2} \frac{f_2f -f_1f_1}{f^2} \d y \\
&= \int -\frac{3f_1^2f_2^2}{f^3} + \frac{3f_1^4f_2}{f^4} \d y \\
&\eqt{\eqref{eqn:app-1241}} \int -\frac{3f_1^2f_2^2}{f^3} + \frac{12f_1^4f_2}{5f^4} \d y .
\end{align*}

\begin{align}
\int \frac{f_{1}f_{2}f_{3}}{f^2} \d y
&= \int \frac{f_{1}f_{2}}{f^2} \d f_{2} \notag \\
&= \int \frac{f_{1}}{2f^2} \d f_{2}^2 \notag \\
&= \frac{f_1f_{2}^2}{2f^2}\bigg |_{y=-\infty}^{+\infty} -\int \frac{f_{2}^2}{2} \left(\frac{f_1}{f^2}\right)_y \d y \notag \\
&\eqt{\eqref{lim:term-zero}}  0 -\int \frac{f_{2}^2}{2}\frac{f_{2}f^2-f_{1}2ff_1}{f^4} \d y \notag \\
&= \int -\frac{f_{2}^3}{2f^2} + \frac{f_1^2f_{2}^2}{f^3} \d y . \label{eqn:app-123111}
\end{align}

\begin{align*}
\int \frac{f_{2}f_{4}}{f}  \d y
&=\int \frac{f_{2}}{f} \d f_{3} \\
&= \frac{f_{2}f_{3}}{f}\bigg |_{y=-\infty}^{+\infty} - \int f_{3}\left(\frac{f_{2}}{f}\right)_y \d y \\
&\eqt{\eqref{lim:term-zero}}  0 - \int f_{3}\frac{f_{3}f-f_{2}f_{1}}{f^2} \d y \\
&=  \int -\frac{f_{3}^2}{f} + \frac{f_{1}f_{2}f_{3}}{f^2} \d y \\
&\eqt{\eqref{eqn:app-123111}} \int -\frac{f_{3}^2}{f} -\frac{f_{2}^3}{2f^2} + \frac{f_1^2f_{2}^2}{f^3}  \d y .
\end{align*}

In the above, the limits are zero due to Proposition~\ref{prop-int-zero}. Since all the integrals and limits exist, all the steps which use integration by parts are valid.
\end{IEEEproof}

\subsection{Proof to Lemma~\ref{lem:deri-4}} \label{app:lem-deri-4}

\begin{IEEEproof}
We use integration by parts to eliminate the high-order terms:
\begin{align}
\int \frac{f_1^6f_2}{f^6}\d y
& = \int \frac{f_1^6}{f^6}\d f_1  \notag \\
& =\int \frac{1}{7f^6}\d f_1^7  \notag \\
& =\frac{f_1^7}{7f^6} \bigg |_{-\infty}^{+\infty} - \int \frac{f_1^7}{7}\left(\frac{1}{f^6}\right)_y \d y \notag \\
& \eqt{\eqref{lim:term-zero}} 0+ \int \frac{6f_1^8}{7f^7} \d y \label{eqn:app-1261}
\end{align}

\begin{align}
\int \frac{f_1^5f_3}{f^5} \d y
& = \int \frac{f_1^5}{f^5}\d f_2 \notag\\
& = \frac{f_1^5f_2}{f^5}\bigg |_{-\infty}^{+\infty}  - \int f_2 \left(\frac{f_1^5}{f^5}\right)_y \d y \notag \\
&\eqt{\eqref{lim:term-zero}} 0 -\int f_2 \frac{5f_1^4}{f^4}\frac{f_2f-f_1^2}{f^2} \d y\notag\\
& =  \int -\frac{5f_1^4f_2^2}{f^5} + \frac{5f_1^6f_2}{f^6} \d y \notag\\
&\eqt{\eqref{eqn:app-1261}} \int -\frac{5f_1^4f_2^2}{f^5} + \frac{30f_1^8}{7f^7} \d y \label{eqn:app-1351}
\end{align}

\begin{align}
\int \frac{f_1^3f_2f_3}{f^4} \d y
& = \int \frac{f_1^3f_2}{f^4} \d f_2 \notag\\
& = \int \frac{f_1^3}{2f^4} \d f_2^2 \notag\\
& = \frac{f_1^3f_2^2}{2f^4}\bigg |_{-\infty}^{+\infty}   -\int \frac{f_2^2}{2}  \left(\frac{f_1^3}{f^4}\right)_y \d y \notag \\
&\eqt{\eqref{lim:term-zero}} 0 -\int\frac{f_2^2}{2}\frac{3f_1^2f_2f^4-f_1^34f^3f_1}{f^8} \d y\notag\\
& =\int -\frac{3f_1^2f_2^3}{2f^4} + \frac{2f_1^4f_2^2}{f^5}\d y \label{eqn:app-123311}
\end{align}

\begin{align}
\int \frac{f_1f_2^2f_3}{f^3} \d y
&= \int \frac{f_1f_2^2}{f^3} \d f_2  \notag\\
&= \int \frac{f_1}{3f^3}\d f_2^3 \notag\\
&= \frac{f_1f_2^3}{3f^3}\bigg |_{-\infty}^{+\infty} -\int \frac{f_2^3}{3}\left(\frac{f_1}{f^3}\right)_y \d y \notag \\
&\eqt{\eqref{lim:term-zero}} 0 - \int \frac{f_2^3}{3}\frac{f_2f^3-f_13f^2f_1}{f^6} \d y \notag\\
&= \int -\frac{f_2^4}{3f^3} +\frac{f_1^2f_2^3}{f^4}\d y \label{eqn:app-123121}
\end{align}

\begin{align}
\int \frac{f_1^4f_4}{f^4} \d y
& =  \int \frac{f_1^4}{f^4} \d f_3 \notag\\
& =\frac{f_1^4f_3}{f^4}\bigg |_{-\infty}^{+\infty} - \int  f_3 \left( \frac{f_1^4}{f^4} \right)_y \d y \notag\\
& \eqt{\eqref{lim:term-zero}} 0 -\int f_3 \frac{4f_1^3}{f^3}\frac{f_2f-f_1^2}{f^2}\d y \notag\\
& = \int -\frac{4f_1^3f_2f_3}{f^4} + \frac{4f_1^5f_3}{f^5}\d y \notag\\
&\eqt{\eqref{eqn:app-123311} \eqref{eqn:app-1351}}
\int -4\left( -\frac{3f_1^2f_2^3}{2f^4} + \frac{2f_1^4f_2^2}{f^5} \right) \notag\\
& \quad + 4 \left( -\frac{5f_1^4f_2^2}{f^5} + \frac{30f_1^8}{7f^7}\right)\d y \notag \\
& =  \int \frac{6f_1^2f_2^3}{f^4} - \frac{28f_1^4f_2^2}{f^5}  + \frac{120f_1^8}{7f^7} \d y \label{eqn:app-1441}
\end{align}

\begin{align}
&  \int \frac{f_1^2f_2f_4}{f^3} \d y \notag \\
& = \int  \frac{f_1^2f_2}{f^3} \d f_3 \notag\\
& = \frac{f_1^2f_2f_3}{f^3}\bigg |_{-\infty}^{+\infty} - \int f_3\left(\frac{f_1^2f_2}{f^3}\right)_y \d y \notag\\
&\eqt{\eqref{lim:term-zero}}  0 -\int f_3 \frac{(f_1^2f_2)_yf^3 - f_1^2f_2(f^3)_y}{f^6} \notag\\
& = \int  -f_3 \frac{(2f_1f_2f_2+ f_1^2f_3)f^3 - f_1^2f_23f^2f_1}{f^6} \d y \notag\\
& =  \int -\frac{2f_1f_2^2f_3}{f^3} - \frac{f_1^2f_3^2}{f^3} +\frac{3f_1^3f_2f_3}{f^4}  \d y \notag\\
&\eqt{\eqref{eqn:app-123121}\eqref{eqn:app-123311}}
\int -2\left(-\frac{f_2^4}{3f^3} +\frac{f_1^2f_2^3}{f^4}\right) - \frac{f_1^2f_3^2}{f^3} \notag \\
& \indent \indent+ 3\left(-\frac{3f_1^2f_2^3}{2f^4} + \frac{2f_1^4f_2^2}{f^5}\right) \d y \notag \\
& =  \int \frac{2f_2^4}{3f^3} - \frac{13f_1^2f_2^3}{2f^4} -\frac{f_1^2f_3^2}{f^3} + \frac{6f_1^4f_2^2}{f^5} \d y \label{eqn:app-124211}
\end{align}

\begin{align}
\int \frac{f_2^2f_4}{f^2} \d y
& = \int \frac{f_2^2}{f^2} \d f_3 \notag \\
& =\frac{f_2^2f_3}{f^2}  \bigg |_{-\infty}^{+\infty}  -  \int  f_3 \left(\frac{f_2^2}{f^2}\right)_y \d y \notag\\
&\eqt{\eqref{lim:term-zero}} 0 -\int f_3 \frac{2f_2}{f}\frac{f_3f-f_2f_1}{f^2} \d y \notag\\
& = \int -\frac{2f_2f_3^2}{f^2} + \frac{2f_1f_2^2f_3}{f^3} \d y \notag\\
&\eqt{\eqref{eqn:app-123121}} \int -\frac{2f_2f_3^2}{f^2} + 2\left(-\frac{f_2^4}{3f^3} +\frac{f_1^2f_2^3}{f^4}\right) \d y \notag\\
& =  \int -\frac{2f_2f_3^2}{f^2} -\frac{2f_2^4}{3f^3} +\frac{2f_1^2f_2^3}{f^4} \d y \label{eqn:app-2421}
\end{align}

\begin{align}
\int \frac{f_1f_3f_4}{f^2} \d y
& = \int \frac{f_1f_3}{f^2} \d f_3 \notag\\
& = \int \frac{f_1}{2f^2} \d f_3^2 \notag\\
& = \frac{f_1f_3^2}{2f^2}\bigg |_{-\infty}^{+\infty}-\int \frac{f_3^2}{2}\left(\frac{f_1}{f^2}\right)_y \d y \notag\\
&\eqt{\eqref{lim:term-zero}}  0 -\int \frac{f_3^2}{2}\frac{f_2f^2-f_12ff_1}{f^4} \d y\notag\\
& =  \int -\frac{f_2f_3^2}{2f^2} + \frac{f_1^2f_3^2}{f^3}\d y\label{eqn:app-134111}
\end{align}

\begin{align}
\int \frac{f_3f_5}{f} \d y
& = \int \frac{f_3}{f} \d f_4 \notag\\
& = \frac{f_3f_4}{f}\bigg |_{-\infty}^{+\infty} -\int f_4 \left(\frac{f_3}{f}\right)_y \d y \notag\\
&\eqt{\eqref{lim:term-zero}}  0 -\int f_4 \frac{f_4f-f_3f_1}{f^2} \d y\notag\\
& = \int -\frac{f_4^2}{f} + \frac{f_1f_3f_4}{f^2} \d y \notag \\
&\eqt{\eqref{eqn:app-134111}} \int -\frac{f_4^2}{f} -\frac{f_2f_3^2}{2f^2} + \frac{f_1^2f_3^2}{f^3} \d y  \label{eqn:app-3511}
\end{align}

\end{IEEEproof}

\ifexpand

\subsection{Proof to Corollary~\ref{col:deri-2}} \label{proof-col1}
\begin{IEEEproof}
Compare \eqref{col-1-1} with \eqref{lemma:1-2},
\begin{align}
&-\frac{1}{2}\int f \left(\frac{f_{yy}}{f}-\frac{f_y^2}{f^2}\right)^2\d y + \frac{1}{2}\int f \left(\frac{f_{yy}}{f}-\frac{f_y^2}{3f^2}\right)^2\d y \notag\\
&= \frac{1}{2} \int \frac{4f_{yy}f_y^2}{3f^2} - \frac{8f_y^4}{9f^3}  \d y \notag \\
&= 0,
\end{align}
where the last step is from \eqref{eq32}.
\end{IEEEproof}

\fi

\ifexpand

\subsection{Proof to Corollary~\ref{col:deri-3}}\label{proof-col2}
\begin{IEEEproof}
When $\frac{1}{3} \leq \beta \leq  \frac{-8+\sqrt{94}}{5}$, the coefficients in  \eqref{col-2-1} are all nonnegative. In fact, $\frac{1}{3} \leq \beta \leq  \frac{-8+\sqrt{94}}{5}$ is the solution of
\begin{align}
             & \frac32-4\beta -\frac54\beta^2 \geq 0, \text{ and }   6\beta - 2 \geq 0.\notag
\end{align}
%\begin{align}
%&\frac{\p^3 }{\p t^3} h(Y_t)\notag\\
%&= \frac12\int f\left( \frac{f_3}{f} - \frac{f_1f_2}{f^2} + \beta \frac{f_1^3}{f^3}\right)^2 + (6\beta - 2) \frac{f_1^2f_2^2}{f^3} + (\frac32-4\beta -\frac54\beta^2) \frac{f_1^4f_2}{f^4}\d y
%\end{align}
%where $\frac{1}{3} \leq \beta \leq  \frac{-8+\sqrt{94}}{5}$.
The proof below is similar to the proof of Theorem~\ref{mainTheorem}.
\begin{align}
 & \frac{1}{2}\int f\left( \frac{f_{3}}{f}-\frac{f_{1}f_{2}}{f^2}+\beta\frac{f_1^3}{f^3}\right)^2   + (6\beta - 2) \frac{f_1^2f_{2}^2}{f^3} + (\frac32-4\beta -\frac54\beta^2) \frac{f_1^4f_{2}}{f^4}  \d y \notag\\
&= \frac{1}{2}\int \frac{f_{3}^2}{f}+  \frac{f_1^2f_{2}^2}{f^3}+ \beta^2\frac{f_{1}^6}{f^5} - \frac{2f_1f_{2}f_{3}}{f^2}  + 2\beta\frac{f_1^3f_{3}}{f^3} \notag \\
&\indent- 2\beta\frac{f_1^4f_{2}}{f^4} +(6\beta - 2) \frac{f_1^2f_{2}^2}{f^3} + (\frac32-4\beta -\frac54\beta^2) \frac{f_1^4f_{2}}{f^4}\d y \notag\\
&= \frac{1}{2}\int \frac{f_{3}^2}{f}+ (6\beta-1) \frac{f_1^2f_{2}^2}{f^3}+ \beta^2\frac{f_{1}^6}{f^5} - \frac{2f_1f_{2}f_{3}}{f^2}  \notag\\
&\indent + 2\beta\frac{f_1^3f_{3}}{f^3} +(\frac32 - 6\beta -\frac54\beta^2) \frac{f_1^4f_{2}}{f^4}\d y \notag\\
&= \frac{1}{2}\int \frac{f_{yyy}^2}{f}+ (6\beta-1) \left(-\frac{f_y^3f_{yyy}}{3f^3} + \frac{f_y^4f_{yy}}{f^4}\right)+ \beta^2\left(\frac{5f_{y}^4f_{yy}}{4f^4}\right)\notag \\
&\indent - \frac{2f_yf_{yy}f_{3}}{f^2}  + 2\beta\frac{f_y^3f_{yyy}}{f^3} +(\frac32 - 6\beta -\frac54\beta^2) \frac{f_y^4f_{yy}}{f^4}\d y \label{tmp1-1}\\
&= \frac{1}{2}\int   \frac{f_{3}^2}{f} + \frac13\frac{f_{1}^3f_{3}}{f^3}  -\frac{2f_1f_{2}f_{3}}{f^2} + \frac12\frac{f_1^4f_{2}}{f^4}   \d y \notag \\
& = ~ \eqref{eq06} \notag
\end{align}
In \eqref{tmp1-1}, the first term enclosed in parentheses is from  \eqref{eq35} and the second is from  \eqref{eq341}.
\end{IEEEproof}

\fi

\section*{Acknowledgment}
The authors would like to thank the Associate Editor and two anonymous reviewers for their comments and suggestions; especially  the second reviewer for providing the references on McKean's Problem and  $\Gamma_2$ calculus.

F. Cheng would like to express his gratitude to Prof. Raymond Yeung for introducing him the topic on EPI when he was a student at CUHK and the support when he was working at the Institute of Network Coding. He would like to thank Prof. Chandra Nair for teaching him the proof in Costa's paper and the help in preparing the manuscript.  He also would like to thank Prof. Venkat Anantharam for his valuable discussion and suggestion, which have greatly improved the paper. He is grateful to Prof. C\'{e}dric Villani for sharing his expertise as well as  the critical comments and providing the references in mathematical physics. The help from Prof. Amir Dembo and Prof. Andrew Barron is sincerely appreciated.

\begin{IEEEbiographynophoto}{\bf Fan Cheng}(S'12-M'14)
received the bachelor degree in
computer science and engineering from Shanghai Jiao Tong
University in 2007, and the PhD degree in
information engineering from The Chinese
University of Hong Kong in 2012. From 2012-2014,
he had been a postdoctoral fellow in the Institute of
Network Coding, The Chinese University of Hong Kong.
As of 2015, he has been a research fellow in the Department of ECE, NUS, Singapore.
\end{IEEEbiographynophoto}

\begin{IEEEbiographynophoto}{\bf Yanlin Geng}(M'12) 
received his B.Sc. (mathematics) and M.Eng. (signal and information processing) from Peking University, and Ph.D. (information engineering) from the Chinese University of Hong Kong in 2006, 2009, and 2012, respectively. He is currently an Assistant Professor in the School of Information Science and Technology, ShanghaiTech University. His research interests are mainly on problems in network information theory.
\end{IEEEbiographynophoto}

\end{document}